\documentclass[reprint,nofootinbib]{revtex4}
\usepackage{graphicx}  
\usepackage{dcolumn}   
\usepackage{bm}        
\usepackage{amssymb}
\usepackage{diagbox}
\usepackage{slashbox}
\usepackage{multirow}
\usepackage{amsmath}
\usepackage{cases}
\usepackage{placeins}
\usepackage{textcomp}
\UseRawInputEncoding

\usepackage{blindtext}
\usepackage{url}
\usepackage{graphicx}
\usepackage{url}
\usepackage{tikz}
\usepackage[bf]{caption}
\usepackage{amsmath}
\usepackage{cases} 
\usepackage{multirow}

\usepackage{epsfig}
\usepackage{makecell}
\usepackage{amsmath}
\usepackage{amsfonts}
\usepackage{amssymb}
\usepackage{mathtools}
\usepackage{url}
\usepackage{subfigure}
\usepackage{hhline}
\usepackage{color}
\usepackage{array,multirow}
\usepackage{bm}
\usepackage{makecell}
\usepackage{epsfig}
\usepackage{amsmath}
\usepackage{amsfonts}
\usepackage{amssymb}
\usepackage{url}
\usepackage{hyperref}
\usepackage{subfigure}
\usepackage{hhline}
\usepackage{color}
\usepackage{bm}
\usepackage{cancel}
\usepackage{mathtools}

\newcommand{\beqa}{\begin{eqnarray}}
\newcommand{\eeqa}{\end{eqnarray}}
\newcommand{\beq}{\begin{equation}}
\newcommand{\eeq}{\end{equation}}

\newcommand{\nn}{\nonumber}
\newcommand{\bmt}{\begin{pmatrix}}
\newcommand{\emt}{\end{pmatrix}}
\usepackage[toc,page]{appendix}
\usepackage{comment}
\newcommand{\be}{\begin{equation}}
\newcommand{\ee}{\end{equation}}
\newcommand{\bea}{\begin{eqnarray}}
\newcommand{\eea}{\end{eqnarray}}

\begin{document}
\title{Combined analysis of $B_c \to D_s^{(*)}\,\mu ^+ \mu ^-$ and $B_c \to D_s^{(*)}\, \nu \bar{\nu}$ decays within $Z^{\prime}$ and leptoquark new physics models}
\author{Manas K. Mohapatra${}^{1}$}
\email{manasmohapatra12@gmail.com} 
\author{N Rajeev${}^{2}$}
\email{rajeev\_rs@phy.nits.ac.in }
\author{Rupak Dutta${}^{2}$}
\email{rupak@phy.nits.ac.in}
\affiliation{
\vspace{0.5cm}${}^1$Department of Physics, Indian Institute of Technology Hyderabad, Kandi - 502285, India\\ 
\vspace{0.5cm}${}^2$National Institute of Technology Silchar, Silchar 788010, India
}

\begin{abstract}
We investigate the exclusive rare semileptonic decays $ B_c \to D_{s}^{(*)}\,(\ell \ell, \nu \bar{\nu})$ induced by neutral current transition 
$b \to s (\ell \ell,\nu \bar{\nu})$ in the presence of non-universal $Z^{\prime}$, scalar and vector leptoquark new physics models. 
We constrain the new physics parameter space by using the latest experimental measurements of $R_{K^{(*)}}$, $P_5^{\prime}$,
$\mathcal{B}(B_s \to \phi \mu^+ \mu^-)$ and $\mathcal{B}(B_s \to  \mu^+ \mu^-)$.
Throughout the analysis, we choose to work with
the particular new physics scenario $C_9 ^{\mu \mu}(NP)=-C_{10} ^{\mu \mu}(NP)$ where both $Z^{\prime}$, $S_{1/3}^3$ and $U_{-2/3}^3$ leptoquarks satisfy the condition.
Using these new coupling parameters we scrutinize the several physical observables such as differential branching fraction, the forward backward asymmetry, 
the lepton polarization asymmetry, the angular observable $P^{\prime}_5$ and the lepton flavor universal sensitive observables including the ratio of branching ratio 
$R_{D_s^{(*)}}$ and the few $Q$ parameters in the $B_c \to D_s^{(*)}\,\mu ^+ \mu ^-$ and $B_c \to D_s^{(*)}\, \nu \bar{\nu}$ decay processes.
\end{abstract}

\maketitle

\section{Introduction}
The hint of new physics (NP) in the form of new interactions which demands an extension of the standard model (SM) of particle physics
are witnessed not only in the flavor changing neutral current decays of rare beauty particles of the form $b \to s\, \ell^+ \ell ^-$ 
but also in the flavor changing
charged current decays proceeding via $b \to c\, \ell\, \nu$ quark level transitions.
The rare weak decays of several composite beauty mesons such as $B_d$, $B_s$, and $B_c$ which are forbidden at the tree level in SM appear to follow loop or box
level diagrams. 
Theoretically, the radiative and semileptonic decays of $B \to K^{(*)}$ and $B_s \to \phi$ processes have received greater attention 
and are studied extensively both within the SM and beyond.
The sensitivity of new physics possibilities in these decays require very good knowledge of the hadronic form factors more specifically for $B \to V$ transitions.
It requires informations from both the light cone sum rule (LCSR) and lattice QCD (LQCD) methods to compute the form factors respectively at low and high $q^2$ regions which eventually confine the whole kinematic region~\cite{Bharucha:2015bzk}.
Currently, we do have the very precise calculations of the form factors that have very accurate SM predictions of the differential branching fractions
and various angular observables in $b \to s \ell ^+ \ell ^-$ decays.
Similarly, the family of neutral decays proceeding via $b \to s \nu \bar{\nu}$ transitions equally provide the interesting opportunity
in probing new physics signatures to that of $b \to s \ell ^+ \ell ^-$ transitions. However so far no experiments have directly addressed any anomalies except the upper bounds of the branching fractions of $B \to K^{(*)} \nu \bar{\nu}$
decay processes. In principle, under the $SU(2)_L$ gauge symmetry both the charged leptons and neutral leptons are treated equally and hence one can extract a close
relation between both $b \to s \ell ^+ \ell ^-$ and $b \to s \nu \bar{\nu}$ decays in beyond the SM scenarios. 
In addition, the decays with $\nu \bar{\nu}$ final state are well motivated for several interesting features since these decays are considered to be theoretically cleaner as they do not suffer from hadronic
uncertainties beyond the form factors such as the non-factorizable corrections and photonic
penguin contributions. 
  
Experimentally, several measurements in $b \to s \ell ^+ \ell ^-$ transitions such as $R_{K^*}=\mathcal{B}(B \to  K^{*}\, \mu^+ \mu ^-)/\mathcal{B}(B \to K^{*} e ^+ e ^-)$ 
from LHCb~\cite{LHCb:2017avl,LHCb:2020lmf} and Belle~\cite{Belle:2019oag} at $q^2\in[0.045, 1.1]$ and $q^2\in[1.1, 6.0]$
show $2.1-2.4\sigma$ deviation from the SM expectations~\cite{Bordone:2016gaq,Hiller:2003js}. 
Similarly, the angular observable $P^{\prime}_5$ in $B \to K^{*} \mu ^+ \mu ^-$ from in the bins
$q^2 \in[4.0, 6.0]$, [4.3, 6.0] and [4.0, 8.0] from ATLAS~\cite{ATLAS:2018gqc}, LHCb~\cite{LHCb:2013ghj,LHCb:2015svh}, CMS~\cite{CMS}, 
Belle~\cite{Belle:2016xuo}
respectively deviate at $3.3\sigma$, $1\sigma$ and $2.1\sigma$ from the SM expectations~\cite{Descotes-Genon:2012isb,Descotes-Genon:2013vna,Descotes-Genon:2014uoa}. 
The recent updates in the measurements of 
$R_K=\mathcal{B}(B \to K \mu ^+ \mu^-)/\mathcal{B}(B \to K e ^+e ^-)$~\cite{LHCb:2021trn,LHCb:2019hip}
in $q^2\in[1.0, 6.0]$ and the branching fraction of $\mathcal{B}(B_s \to \phi \mu ^+ \mu ^-)$~\cite{LHCb:2021zwz,LHCb:2013tgx,LHCb:2015wdu} 
in $q^2\in[1.1, 6.0]$ region from LHCb still indicate 
$3.1\sigma$ in $R_K$~\cite{Bordone:2016gaq,Hiller:2003js} and $3.6\sigma$ in $\mathcal{B}(B_s \to \phi \mu ^+ \mu ^-)$~\cite{Aebischer:2018iyb,Bharucha:2015bzk} from the SM expectations. 
Similarly, the measurements pertaining to $b \to s \nu \bar{\nu}$ transitions, the upper bound measured by the Belle collaboration in the branching fraction of $B \to K^{(*)} \nu \bar{\nu}$ decays are
$\mathcal{B}(B \to K \nu \bar{\nu})<1.6\times 10^{-5}$ and $\mathcal{B}(B \to K^* \nu \bar{\nu})< 2.7\times 10^{-5}$~\cite{Belle:2017oht} respectively. 
There also exist the BaBar measurement on $\mathcal{B}(B \to K^* \nu \bar{\nu}) < 4\times 10^{-5}$~\cite{BaBar:2013npw}.
Very recently, the Belle II updated the upper bound of $\mathcal{B}(B \to K \nu \bar{\nu}) < 4.1\times 10^{-5}$ in 2021~\cite{Browder:2021hbl}.

There exist several other decay channels similar to the $B \to K^{(*)}$ and $B_s \to \phi$ undergoing 
$b \to s \ell ^+ \ell ^-$ quark level transitions~\cite{Li:2011nf,Huang:2018rys,Falahati:2014yba,Ahmed:2010tt,Capdevila:2017bsm,Bashiry:2009wq,Faustov:2014zva,Wang:2012ab,Li:2010ra,Rajeev:2020aut,Browder:2021hbl,Bobeth:2001jm,Altmannshofer:2009ma,Descotes-Genon:2020buf,Fajfer:2018bfj,Li:2018lxi,Alok:2019xub}.
If any new physics present in $B \to K^{(*)}$ and $B_s \to \phi$ decays can in principle be reflected in several other decays as well.
In that sense we choose to study the explicit rare decays $B_c \to D_s ^{(*)} \mu^+ \mu^-$ and $B_c \to D_s ^{(*)} \nu \bar{\nu}$ which undergo similar $b \to s$ neutral transition.
The particular decay modes $B_c \to D_s ^{(*)} \mu^+ \mu^-$ have been studied previously in SM using various form factors which include relativistic quark model (RQM)~\cite{Ebert:2010dv}, the light front and constituent quark model~\cite{Geng:2001vy,Choi:2010ha}, the three point QCD sum rules approach~\cite{Azizi:2008vv}, the covariant quark model~\cite{Issadykov:2017wlb} and very recently using the lattice QCD form factors only for $B_c \to D_s$ transitions~\cite{Cooper:2021ofu}. In addition,
as far as beyond SM analysis are concerned these decays have also been analyzed within model independent and dependent new physics as well. 
The model independent study within the effective field theory approach was done in Ref.~\cite{Dutta:2019wxo} under various 1D and 2D NP scenarios.
In Ref.~\cite{Lu:2012qnh}, the decay mode was studied within non-universal $Z ^{\prime}$ model with the NP contribution coming from only the right handed currents.
Similarly, the contribution of $Z ^{\prime}$ was analysed by considering the UTfit inputs of left-handed coupling of $Z ^{\prime}$ boson and the different values of
new weak phase angle in the Ref.~\cite{Maji:2020zlq}. 
Similarly, the SM results pertaining to the $B_c \to D_s ^{(*)} \nu \bar{\nu}$ decays have been addressed in Refs.~\cite{Ebert:2010dv,Choi:2010ha,Wang:2014yia}. 

Even from the experimental point of view after the discovery of $B_c$ meson at CDF via $B_c \to J/ \Psi \ell \nu$~\cite{CDF:1998ihx}, the study of $B_c$ decays were found to be very interesting. Unlike the weak decays of other $B$ mesons, the $B_c$ mesons are interesting as it is composed of both heavy $b$ and $c$ quarks that allows broader kinematic range which eventually involve large number of decays.
In addition, the upcoming LHC run can produce around $10 ^8 - 10 ^{10}$ $B_c$ mesons~\cite{Du:1988ws, Chang:1992jb, Cheung:1993qi, Braaten:1993jn, Stone:1997vk} which offer very rich laboratory for the associated $B_c$ weak decays. At LHC with luminosity $10^{34}$ $cm^{-2}s^{-1}$, one could expect around $2 \times 10^{10}$ $B_c$ events per year~\cite{Gouz:2002kk, PepeAltarelli:2008yyl}. 
Hence, exploring the new physics in $B_c \to D_s^{(*)} \mu ^+ \mu ^-$ decays are experimentally well motivated at the LHCb experiments.

For the experimental scope of $B_c \to D_s ^{(*)} \nu \bar{\nu}$ channels are concerned,
as we know, they are experimentally challenging because of the di-neutrinos in the final state which leave no information in the detector. 
However, recently Belle II at SuperKEK uses a novel and independent inclusive tagging approach and measures the 
upper limit of the branching fraction of $B \to K \nu \bar{\nu}$ decays~\cite{Belle-II:2021rof}. This novel method has benefited 
with larger signal efficiency of about 4\%, at the cost of higher background level~\cite{Belle-II:2021rof}. 
Since Belle and Belle II mainly work at $\Upsilon (4S)$ and $\Upsilon (5S)$ resonances, no $B_c$ mesons are produced. 
Hence study of $B_c \to D_s ^{(*)} \nu \bar{\nu}$ channels would be difficult at Belle II experiment.
Moreover, as there are more number of $B_c$ mesons produced at LHC, can in principle, 
would be feasible to predict an upper limit of the branching fraction of $B_c \to D_s ^{(*)} \nu \bar{\nu}$ decays in their future prospects.


In this context, we study the implication of the latest $b \to s \ell ^+\ell ^-$ data on the $B_c \to D_s ^{(*)}\mu ^+ \mu ^-$ and $B_c \to D_s ^{(*)}\nu \bar{\nu}$ 
decay processes under the model dependent analysis. 
We choose in particular, the specific models such as $Z ^{\prime}$ and the various scalar and vector leptoquarks (LQs) which satisfy $C_9 ^{\mu \mu}(NP)=-C_{10} ^{\mu \mu}(NP)$ new physics scenario.
Among various LQs, we opt for the specific LQs in such a way that they should have combined NP effects in the form of $C_9 ^{\mu \mu}(NP)=-C_{10} ^{\mu \mu}(NP)$ both in $b \to s \ell ^+ \ell ^-$
and $b \to s \nu \bar{\nu}$ decays. 
Hence, the main aim of this work is to extract the common new physics that appear simultaneously in $b \to s \ell ^+\ell ^-$ and
$b \to s \nu \bar{\nu}$ decays.

The layout of the present paper is as follows. In Section \ref{Thfm}, we add a theoretical framework that includes a brief discussion of effective Hamiltonian for $b\to s \ell^+ \ell^-$ and $b \to s \nu \bar{\nu}$ parton level transition. In addition to this, we also present the differential decay distributions and other $q^2$ dependent observables of $B_c \to D_s^{(*)} \mu ^+ \mu^-$ and $B_c \to D_s^{(*)} \nu \bar{\nu}$ processes. In the context of new physics, we deal with the contributions arising due to the exchange of $LQ$ and $Z^{\prime}$ particles in Section \ref{NPanalysis}. In Section \ref{Num_analysis}, we report and discuss our numerical analysis in the SM and in the presence of NP contributions. Finally we end with our conclusion in Section \ref{conc}.

\section{Theoretical Framework}\label{Thfm}
\subsection{Effective Hamiltonian}
The effective Hamiltonian responsible for $b \to s \ell \ell$ parton level transition in the presence of NP vector operator can be represented as~\cite{Buras:1994dj}
\bea
&&{\cal H}_{\rm eff}=-\frac{\alpha G_F}{\sqrt 2 \pi} V_{tb}V_{ts}^* \Big[2 \frac{C_7^{\rm eff}}{q^2} \left [\bar s \sigma^{\mu \nu} q_\nu (m_s P_L +m_b P_R) b \right ]  (\bar \ell \gamma_\mu \ell)+ C_9^{\rm eff} (\bar s \gamma^\mu P_L b) (\bar \ell \gamma_\mu \ell) \nn\\
&&~~~~~~+ C_{10} (\bar s \gamma^\mu P_L b) (\bar \ell \gamma_\mu \gamma_5 \ell)+{C}_{9}^{\ell \ell} (\rm NP)\, (\bar{s}\, \gamma^{\mu}\, P_L\, b\,) (\bar{\ell}\, \gamma_{\mu}\, \ell\,) +{C}_{10}^{\ell \ell} (\rm NP)\, \bar{s}\, \gamma^{\mu}\, P_L\, b\, \bar{\ell}\, \gamma_{\mu}\, \gamma_{5}\, \ell\,\Big],\label{Ham-SM}
\eea
where $G_F$ is the Fermi coupling constant, $\alpha$ is the fine structure constant, $V_{ij}$ is the CKM matrix element, and $C_7 ^ {\rm eff}, C_9 ^ { \rm eff}$ and $C_{10}$ are the relevant Wilson coefficients (WC) evaluated at $\mu =m_b^ {\rm pole}$ scale \cite{Buras:1994dj}. The Wilson coefficients $C_9^{\ell \ell} (\rm NP)$ and $C_{10}^{\ell \ell} (\rm NP)$ are the effective coupling constants associated with the corresponding NP operators.
The effective Hamiltonian describing $b \to s \nu \bar{\nu}$ decay processes is given by \cite{Melikhov:1997wp}
\bea \label{Ham_nunubar}
\mathcal{H}_{\rm eff}^ {\nu \bar{\nu}}= \frac{G_F \alpha}{2 \sqrt{2}\pi} V_{tb} V_{ts}^* C_L^{\nu \nu} \mathcal{O}_L^{\nu \nu}.
\eea
Here the effective four fermion operator $\mathcal{O}_L^{\nu \nu}= (\bar{s} \gamma_\mu P_L b) (\bar{\nu} \gamma^\mu (1- \gamma_5) \nu)$ and the associated coupling strength $C_L^{\nu \nu} = X(x_t)/s_W^2$  which includes the Inami - Lim function $X(x_t)$ given in Ref.~\cite{Buchalla:1995vs}.
In principle, several Lorentz structures in the form of chiral operators can be possible in the NP scenario such as vector, axial vector, scalar, pseudoscalar, and tensor. However, among all the operators scalar, pseudoscalar and tensor are severely constrained by $B_s \to \mu \mu$ and $b \to s \gamma$ measurements~\cite{Bardhan:2017xcc}. Therefore we consider the vector and axial vector contributions only.
In our analysis, among possible NP operators we consider only the left chiral $\mathcal{O}_9 ^{\ell \ell} (\rm NP)$ and $\mathcal{O}_{10}^ {\ell \ell} (\rm NP)$ contributions and the associated Wilson coefficients are assumed to be real.
The effective Wilson coefficients $C_7^{\rm eff}$ and $C_9^{\rm eff}$ are defined as \cite{Ali:1999mm}
\bea
 {C}_{7}^{eff} &=& {C}_7 - \frac{{C}_5}{3} - {C}_6 \nn \\ 
 {C}_{9}^{eff} &=& {C}_{9}(\mu)\,+\,h(\hat{m}_c, \hat{s})\,{C}_{0}\, -\,\frac{1}{2}\,h(1,\hat{s})
 (4 {C}_{3}\,+\,4 {C}_{4}\,+\,3 {C}_{5}\,+\,{C}_{6})\, \nonumber \\ &&
 -\, \frac{1}{2}\,h(0,\hat{s}) ({C}_{3}\,+\, 3 {C}_{4})\, +\, \frac{2}{9}(3 {C}_{3}\,+ {C}_{4}\,+\,3 {C}_{5}\,+\,{C}_{6})\,,
\eea
where $\hat{s}=q^2/m_{b}^2$, $\hat{m}_c=m_c/m_b$ and
${C}_0=3 {C}_1\, +\, {C}_2\, +\, 3{C}_3\, +\,{C}_4\, +\, 3{C}_5\, +\, {C}_6$.
\bea
 h(z,\hat{s})=-\frac{8}{9}\ln \frac{m_b}{\mu}-\frac{8}{9}\ln z + \frac{8}{27} + \frac{4}{9}x-\frac{2}{9}(2+x)
 |1-x|^{1/2}
 \begin{cases}
\ln \lvert \frac{\sqrt{1-x}+1}{\sqrt{1-x}-1}\rvert -i \pi\,, & \text{for $x \equiv \frac{4z^2}{\hat{s}}<1$} \\
             2 \arctan \frac{1}{\sqrt{x-1}}, & \text{for $x \equiv \frac{4z^2}{\hat{s}}>1$}
\end{cases}
\eea
and
\bea
 h(0,\hat{s})=-\frac{8}{9}\ln \frac{m_b}{\mu}\,-\,\frac{4}{9}\ln \hat{s}\, +\, \frac{8}{27}\, +\, \frac{4}{9} i \pi.
\eea 

Here we have included the short distance perturbative contributions to $C_9 ^{\rm eff}$.
The measurements of the $B\to( K^{(*)}, \phi)\ell \ell$ processes induced by $b \to s \ell \ell$ quark level transition, in principle, include the available vector resonances. 
These regions arise in the dimuon invariant mass resonance $m_{\mu \mu}$ around $\phi (1020), J/ \psi (3096)$ and $\psi_{2S}(3686)$ along with broad charmonium states $\psi (3770), \psi (4040), \psi (4160)$ and $\psi(4415)$. The amplitudes in these regions though are dominated but have a large theoretical uncertainty. However this is performed with a model describing these vector resonances as a sum of the relativistic Breit-Wigner amplitudes.
The explicit expressions of the $c\bar{c}$ resonance part reads
\bea
\mathcal{Y}_{BW}(q^2)=\frac{3\pi}{\alpha ^2}\sum _{V_i=J/\psi, \psi^{\prime}}\frac{\Gamma (V_i \to \ell ^+ \ell ^-)M_{V_i}}{M_{V_i}^2-q^2-iM_{V_i}\Gamma _{V_i}},
\eea
where $M_{V_i}(J/\psi, \psi^{\prime})$ is the mass of the vector resonance and $\Gamma _{V_i}$ is the total decay width of the vector mesons. Moreover, in our study we have taken care by avoiding all 
the $c\bar{c}$ resonances appearing at respective $q^2$ regions. Hence, our all predictions concerned to $b \to s \mu ^+ \mu ^-$ decay observables are done in the $q^2$ regions
[0.1 - 0.98] and [1.1 - 6] $\rm GeV^2$.
On the other hand, the case with $b \to s \nu \bar{\nu}$ quark level transitions are quite different. In principle, $b \to s \nu \bar{\nu}$ decays are free from various
hadronic uncertainties beyond the form factors such as the non-factorizable corrections and photonic penguin contributions.
In particular, the vector charmonium states including $\phi (1020), J/ \psi (3096)$ and $\psi_{2S}(3686)$ etc do not contribute to any di-neutrino states
in the concerned decays of $b \to s \nu \bar{\nu}$ channels. 
Hence one can access the whole $q^2$ region in the prediction of various observables in $B_c\to D_s^{(*)}\nu \bar{\nu}$ decays. 
Hence, the $b \to s \nu \bar{\nu}$ decay channels are treated as theoretically cleaner than the $b \to s \ell \ell$ decay processes.
Apart from this, additionally we do not even include any non local effects in our analysis which are important below the charmonium contributions that have been studied in detail in Refs.~\cite{Khodjamirian:2010vf,Khodjamirian:2012rm,Bobeth:2017vxj,Gubernari:2020eft}. In principle, these hadronic non local effects are generally neglected in the study of the lepton flavor universality violation (LFUV) in various $b \to s \ell \ell$ decays. 
In addition, the factorizable effects arising due to the spectator scattering may not affect severely to the LFU ratios and hence these effects are neglected in our present analysis~\cite{Beneke:2001at,Altmannshofer:2008dz}.
\subsection{Differential decay distribution and $q^2$ observables in $B_c \to D_s^{(*)} \mu^+ \mu ^-$}

In analogy with $B\to K\ell^+\ell^-$ decay mode, it is useful to note that the rare semileptonic $B_c \to D_s \ell^+ \ell^-$ process is also mediated through $b\to s \ell^+ \ell^-$ transition in the parton level. In the standard model, we present the formula of $q^2$ dependent differential branching ratio which is given as follows \cite{Bouchard:2013eph}:
\begin{equation}
\frac{dBR}{dq^2} = \frac{\tau_{B_c}}{\hbar}(2a_\ell + \frac{2}{3}c_\ell) ,
\end{equation}
where the parameters $a_\ell$ and $c_\ell$ are given by
\begin{eqnarray}
a_\ell &=& \frac{G_F^2 \alpha_{EW}^2 |V_{tb} V_{ts}^*|^2}{2^9\pi^5 m_{B_c}^3}\beta_\ell \sqrt{\lambda} \Big[ q^2 |F_P|^2 + \frac{\lambda}{4}(|F_A|^2 + |F_V|^2) + 4m_\ell^2 m_{B_c}^2 |F_A|^2 \nonumber \\
&+& 2m_\ell(m_{B_c}^2 - m_{D_s}^2 + q^2){\rm Re}(F_PF_A^*) \Big], \\
c_\ell &=& -\frac{G_F^2 \alpha_{EW}^2 |V_{tb} V_{ts}^*|^2}{2^9\pi^5 m_{B_c}^3}\beta_\ell \sqrt{\lambda}\frac{ \lambda \beta_\ell^2}{4}(|F_A|^2 + |F_V|^2).
\end{eqnarray}
Here the kinematical factor $\lambda$ and the mass correction factor $\beta_\ell$ in the above equations are given by
\begin{eqnarray} \label{lambda}
\lambda &=& q^4 + m_{B_c}^4 + m_{D_s}^4 - 2(m_{B_c}^2m_{D_s}^2 + m_{B_c}^2q^2 + m_{D_s}^2q^2), \nn\\
\beta_\ell &=& \sqrt{1-4m_\ell^2/q^2}.
\end{eqnarray}
However, the explicit expressions of the form factors such as $F_P$, $F_V$ and $F_A$ are given as follows:
\begin{eqnarray}
F_P &=& -m_\ell C_{10} \Big[ f_+ - \frac{m_{B_c}^2-m_{D_s}^2}{q^2}(f_0-f_+) \Big], \\
F_V &=& C_9^{\rm eff} f_+ + \frac{2m_b}{m_{B_c}+m_{D_s}}C_7^{\rm eff}f_T, \\
F_A &=& C_{10}f_+.
\end{eqnarray}
We employ the Wilson coefficients at the renormalization scale $\mu = 4.8$ GeV as reported in Ref.~\cite{Bouchard:2013eph}.

Similarly, the transition amplitude for $B_c\to D_s^*\ell^+\ell^-$ decay channel can be obtained from the effective Hamiltonian given in the Eq. (\ref{Ham-SM}). 
The $q^2$ dependent differential branching ratio for $B_c\to D_s ^* \ell^+\ell^-$ process is given as~\cite{Descotes-Genon:2015uva}
\bea
d\mathcal{BR}/dq^2=\frac{d\Gamma / dq^2}{\Gamma _{Total}}=\frac{\tau _{B_c}}{\hbar} \frac{1}{4}\bigg[3I_{1}^c + 6I_{1}^s - I_{2}^c -2I_{2}^s\bigg],
\eea
where the $q^2$ dependent angular coefficients are given in Appendix~\ref{Ang_coeff} and the corresponding SM WCs at the renormalization scale $\mu = 4.8$ GeV are taken from~\cite{Ali:1999mm}.
In addition to this, we also define other prominent observables such as the forward-backward asymmetry $A_{FB}$, the longitudinal polarization fraction $F_L$
and the angular observable $P'_5$ which are given by~\cite{Descotes-Genon:2012isb}
\begin{equation}
 F_L(q^2)=\frac{3I_{1}^c - I_{2}^c}{3I_{1}^c + 6I_{1}^s - I_{2}^c -2I_{2}^s}, \hspace{0.1cm}
 A_{FB}(q^2)=\frac{3I_{6}}{3I_{1}^c + 6I_{1}^s - I_{2}^c -2I_{2}^s}, \nonumber
\end{equation}
\begin{equation}
\langle P^{\prime}_5 \rangle = \frac{\int_{bin}dq^2 I_{5}}{2 \sqrt{-\int_{bin}dq^2 I_{2}^c \int_{bin}dq^2 I_{2}^s}}.
\end{equation}

To confirm the existence of the lepton universality violation,  one can construct additional observables associated with the two different families of lepton pair which are quite sensitive to shed light into the windows of NP. The explicit expressions are given as below~\cite{Descotes-Genon:2013vna,Capdevila:2016ivx}:
\begin{equation}
 \langle Q_{F_L} \rangle = \langle {F^ \mu_L} \rangle - \langle {F^e_L} \rangle, \hspace{0.5cm}
 \langle Q_{A_{FB}} \rangle = \langle {A^ \mu_{FB}} \rangle - \langle {A^e_{FB}} \rangle, \hspace{0.5cm}
 \langle Q_{5}^{\prime} \rangle = \langle P_{5}^{\prime\mu} \rangle - \langle P_{5}^{\prime e} \rangle.
\end{equation}

Also we define the ratio of the branching ratios of $\mu$ to $e$ transition in $B_c\to D_s^{(*)} \ell^ + \ell^-$ decay modes as follows:
\begin{eqnarray}
R_{D_s^{(*)}}(q^2) = \frac{\mathcal{BR}\,\Big(B_c \to D_s^{(*)}\, \mu^+\mu^-\Big)}
{\mathcal{BR}\,\Big(B_c \to D_s^{(*)}\, e^+\,e^-\Big)}\,.
\end{eqnarray}

\subsection{Differential decay distribution in $B_c \to D_s^{(*)} \nu \bar{\nu}$}
The explicit study of $B_c\to D_s^{(*)} \nu \bar{\nu}$ processes involved with $b\to s\nu \bar{\nu}$ transitions are also quite important to search for NP beyond the SM as they are associated to $b\to s \ell^+ \ell^-$ parton level by $SU(2)_L$ symmetry group.
From the effective Hamiltonian given in Eq. \ref{Ham_nunubar}, the differential decay rate for $B_c \to D_s^{(*)} \nu \bar{\nu}$ decay channel is given by~\cite{Altmannshofer:2009ma,Buras:2014fpa}
\begin{align}
\frac{d\mathcal{BR}(B_c\to D_s\nu\bar\nu)_\text{SM}}{dq^2}
&= \tau_{B_c} 3|N|^2\frac{X_t^2}{s_w^4} \rho_{D_s}(q^2),
\\
\frac{d\mathcal{BR}(B_c\to D_s ^*\nu\bar\nu)_\text{SM}}{dq^2}
&= \tau_{B_c}3|N|^2\frac{X_t^2}{s_w^4}  \left[\rho_{A_{1}}(q^2)+\rho_{A_{12}}(q^2)+\rho_V(q^2)\right],
\label{eq:FLSM}
\end{align}
where the factor 3 comes from the sum over neutrino flavors, and 
\begin{equation}
N= V_{tb}V_{ts}^*\,
\frac{G_F \alpha}{16\pi^2 }
\sqrt{\frac{m_{B_c}}{3\pi}}\,
\end{equation}
is the normalization factor. The relevant rescaled form factors $\rho_i$ given in the above equations are given below.
\begin{align}
\rho_{{D_s}}(q^2)&=
\frac{\lambda^{3/2}_{D_s}(q^2)}{m_{B_c}^4}   \left[f^{D_s}_+(q^2)\right]^2, \hspace{3.2cm} \rho_V(q^2)=
\frac{2 q^2 \lambda^{3/2}_{{D_s}^*}(q^2)}{(m_{B_c} + {m}_{{D_s}^*})^2m_{B_c}^4}\left[V(q^2)\right]^2, \nn \\
\rho_{A_{1}}(q^2)&=
 \frac{2q^2\lambda^{1/2}_{{D_s}^*}(q^2)(m_B + {m}_{{D_s}^*})^2}{m_{B_c}^4}\left[A_1(q^2)\right]^2, \hspace{0.3cm} \rho_{A_{12}}(q^2)=
\frac{64   {m}_{{D_s}^*}^2 \lambda^{1/2}_{{D_s}^*}(q^2)}{m_{B_c}^2}  \left[A_{12}(q^2)\right]^2.
\end{align}
The parameter
$\lambda$ is already defined for $B_c \to D_s$ transition in Eq.~(\ref{lambda}) and the pseudoscalar $D_s$ is replaced by the vector meson $D_s^*$ in $B_c \to D_s^*$ decays.

\section{New Physics Analysis in the scenario $C_9 ^{\mu \mu}(\rm NP)=-C_{10} ^{\mu \mu}(\rm NP)$} \label{NPanalysis}

Assuming the NP exist only in the context of $\mu$ mode in $b\to s \ell^+\ell^-$ transition, it will contribute to more number of Lorentz structures. 
The new physics scenarios for the parton level $b\to s \mu ^+ \mu ^-$ transition that account for NP contributions are given as~\cite{Descotes-Genon:2015uva,Capdevila:2017bsm} 
\bea \label{WC_scenario}
(\rm I):&&C_9^ {\mu \mu}(\rm NP)< 0, \nn\\
(\rm II):&&C_9^ {\mu \mu}(\rm NP) = -C_{10}^ {\mu \mu}(\rm NP)< 0, \nn \\
(\rm III):&&C_9^ {\mu \mu}(\rm NP) =-C_{9}^{\prime \mu \mu}(\rm NP)< 0, \nn\\
(\rm IV):&&C_9^{\mu \mu}(\rm NP)=-C_{10}^{\mu \mu}(\rm NP)=C_9^{\prime \mu \mu}(NP)=C_{10}^{\prime \mu \mu}(NP)< 0,
\eea
where the unprimed couplings differ from the primed Wilson coefficients by their corresponding chiral operator as discussed in the previous section. Keeping in mind, as from the Ref.~\cite{Alok:2017jgr}, only three out of ten leptoquarks such as $S_3, U_1$ and $U_3$ can explain the $b\to s \mu^+ \mu^-$ data as they have good fits under certain scenario. Among  $S_3$, $U_3$ and $U_1$ leptoquarks, the $U_1$ leptoquark  has no contribution to the couplings corresponding to the NP operator responsible for $b\to s \nu \bar{\nu}$ processes whereas other two LQs are differentiated with a definite contributions to it. Hence the effect from $U_1$ LQ is not taken into account in the present analysis. It is important to say that the remaining $S_3$ and $U_3$ LQs do not satisfy the scenarios I and III. On the other hand, in Ref. \cite{Alok:2017sui} it is also reported that $Z^{\prime}$ can contribute to both scenario I and II whereas a vast majority of this model use the scenario II. Hence the feasible environment to study  both LQs and $Z^ {\prime}$ simultaneously will be the scenario $\rm II:C_9^ {\mu \mu}(NP) = -C_{10}^ {\mu \mu}(NP)$.
Many works have been studied in these scenarios in LQs \cite{Calibbi:2015kma, Alonso:2015sja, Hiller:2014yaa, Gripaios:2014tna, deMedeirosVarzielas:2015yxm, Sahoo:2015wya, Fajfer:2015ycq, Becirevic:2015asa, Becirevic:2016oho} and in the presence of $Z'$~\cite{Calibbi:2015kma, Greljo:2015mma, Chiang:2016qov,Gauld:2013qba, Gauld:2013qja, Crivellin:2015era,Ahmed:2017vsr}. Therefore, the purpose of this work is to concentrate on the scenario II : $C_9 ^{\mu \mu}(\rm NP)=-C_{10} ^{\mu \mu}(\rm NP)$~\cite{Alok:2017sui}.

\subsection{Leptoquark contribution}
There are 10 different leptoquark multiplets under the SM gauge group $SU(3)_C \times SU(2)_L\times U(1)_Y$ in the presence of dimension $\leq 4$ operators \cite{Buchmuller:1986zs} in which
five multiplets include scalar (spin 0) and other halves are vectorial (spin 1) in nature under the Lorentz transformation. 
Among all, both the scalar triplet $S^3$ (Y= 1/3) and vector isotriplet $U^3$ (Y= -2/3) can explain the $b \to s \mu^ + \mu ^-$ and $ b \to s \nu \bar{\nu}$ processes simultaneously. 
The relevant Lagrangian is given as follows \cite{Alok:2017jgr}
\bea
\mathcal{L}_S &=& y_{\ell q}' \bar{\ell^c_L} i \tau _2 \overrightarrow{\tau} q_L S^3_{1/3} + h.c, \nn\\
\mathcal{L}_V &=& g_{\ell q}' \bar{\ell}_L \gamma _\mu \overrightarrow{\tau} q_L U^3_{-2/3} + h.c,
\eea
where the fermion currents in the above Lagrangian include the $SU(2)_L$  quark and lepton doublets ``$q_L$" and $``\ell _L"$ respectively, and $\tau$ represent the Pauli matrices. Most importantly the parameters $y_{\ell q}^{\prime}$ and $g_{\ell q}^{\prime}$ are the quark - lepton couplings associated with the corresponding leptoquarks. In particular for this analysis ${y'}_{\ell q} ^{\mu b(s)}$ is the coupling of the leptoquark $S_{1/3}^3$ to the left-handed $\mu$ or {$\nu _\mu$}and a left-handed fermion field $b$ $(s)$. Similarly ${g'}_{\ell q} ^{\mu b(s)}$ is the coupling correspond to the leptoquark $U_{-2/3}^3$. On the other hand,   for the parton level $b\to s \nu \bar{\nu}$ transitions both $S_{1/3}^3$ and $U_{-2/3}^3$  LQs contribute differently as reported in Table \ref{LQWC}. Hence the Wilson coefficient $C_L^{\nu \nu} $ associated with $b\to s \nu \bar{\nu}$ can be obtained by replacing $C_L^{\nu \nu} \to C_L^{\nu \nu} + C_L^{\nu \nu} (\rm NP)$. In our paper, we consider the couplings $y_{\ell q}^{\prime \mu b} (y_{\ell q}^{\prime \mu s})^*$ and $g_{\ell q}^{\prime \mu b} (g_{\ell q}^{\prime \mu s})^*$ as real for the $S_{1/3}^3$ and $U_{-2/3}^3$ LQs respectively with the assumption of same mass for both the leptoquarks.

\begin{table}
\centering
\setlength{\tabcolsep}{8pt} 
\renewcommand{\arraystretch}{1.5} 
\begin{tabular}{|c|ccc|} \hline
NP model & $C_9^{\mu\mu}({\rm NP})$ & $C_{10}^{\mu\mu}({\rm NP})$ & $C_L^{\nu\nu}({\rm NP})$ \\
\hline
\hline
$S^3_{1/3}$ & $\mathcal{R} y_{\ell q}^{\prime \mu b} (y_{\ell q}^{\prime \mu s})^*$ & $-\mathcal{R} y_{\ell q}^{\prime \mu b} (y_{\ell q}^{\prime \mu s})^*$ & $\frac12 \mathcal{R}  y_{\ell q}^{\prime \mu b} (y_{\ell q}^{\prime \mu s})^*$ \\
\hline
$U^3_{-2/3}$ & $-\mathcal{R} g_{\ell q}^{\prime \mu b} (g_{\ell q}^{\prime \mu s})^*$
                                   & $\mathcal{R} g_{\ell q}^{\prime \mu b} (g_{\ell q}^{\prime \mu s})^*$ & $-2\mathcal{R} g_{\ell q}^{\prime \mu b} (g_{\ell q}^{\prime \mu s})^*$\\
\hline
\hline
$Z'$ & $-\mathcal{M} g_{L}^{bs} g_{L}^{\mu \mu}$
 & $\mathcal{M} g_{L}^{bs} g_{L}^{\mu \mu}$ & $-\mathcal{M} g_{L}^{bs} g_{L}^{\nu \bar{\nu}}$\\
\hline
\end{tabular}
\caption{\large{Contributions of the LQs - $S^3_{1/3}, U^3_{-2/3}$, and $Z'$ to the Wilson coefficients. The normalization $\mathcal{R(M)} \equiv \pi / (\sqrt{2}
  \alpha G_F V_{tb} V_{ts}^* (M_{LQ}(M_{Z'}))^2$ and $M_{LQ} = M_{Z'} = 1$ TeV}.
\label{LQWC}}
\end{table}
\subsection{Non-universal $Z'$ contribution}
The extension of SM by an extra minimal $U(1)'$ gauge symmetry produces a neutral gauge boson the so called $Z'$ boson. It is the most obvious candidate which represent to $b \to s \mu ^+ \mu ^-$ in the NP scenario. However the main attraction of this model includes the flavor changing neutral current (FCNC) transition in the presence of new non-universal gauge boson $Z'$ \cite{Langacker:2000ju, Barger:2003hg, Barger:2004hn} which can contribute at tree level. After integrating out the heavy $Z'$ the effective Lagrangian for 4 fermion operator is given as

\begin{figure}[htbp]
\centering
\includegraphics[scale=0.4]{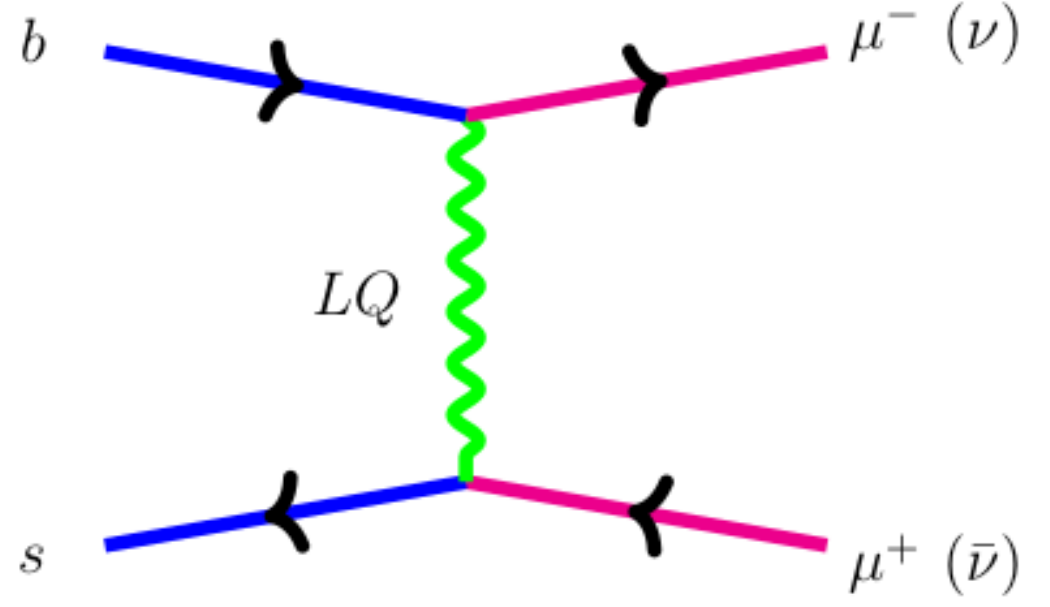}
\includegraphics[scale=0.4]{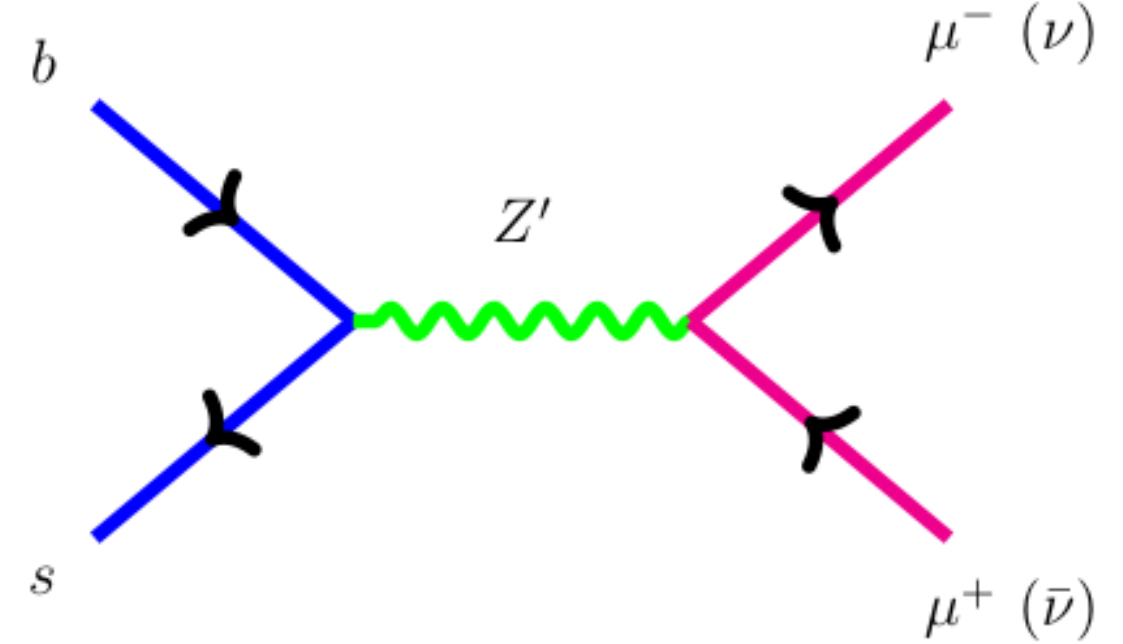} 
\caption{\large{The tree level contribution of the LQs and the $Z'$ for $B_c \to D_s^{(*)}(\mu ^+ \mu ^-, \nu \bar{\nu})$}}
\label{feyn}
\end{figure}

\bea
\mathcal{L}_{Z'}^{eff} = - \frac{1}{2 M_{Z'}^2}J_ \mu ^ \prime  J ^{ \mu \prime},
\eea
where the new current is given as 
\bea
J^{ \mu \prime} = - g^{\mu \mu} _ {LL} \bar{L}\gamma ^ \mu P_L L + g^{\mu \mu} _ {L(R)} \bar{\mu}\gamma ^ \mu P_{L(R)} \mu + g_L ^ {ij} \bar{\psi} _i \gamma ^ \mu P_L \psi _j + h.c.
\eea
where $i$ and $j$ are family index, $P_{L(R)}$ is the projection operator of left (right) chiral fermions and $g_L^{ij} $ denote the left chiral coupling of $Z'$  gauge boson.
Now relevant interaction Lagrangian is given as 
\bea
\mathcal{L} _{Z'}^{eff}= -\frac{g_L^{bs}}{m_{Z'}^2}(\bar{s} \gamma ^ \mu b) (\bar{\mu} \gamma ^ \mu (g_L ^{\mu \mu}P_L + g_R ^{\mu \mu} P_R)\mu).
\eea
Now, the modified Wilson coefficients in the presence of $Z'$ model can be written as \cite{Huang:2018rys}
\bea
C_9^{\mu \mu}(\rm NP) &=& - \bigg[\frac{\pi}{\sqrt{2}G_F \alpha V_{tb}V_{ts}^*}\bigg]\frac{g_L^{bs}(g_L^ {\mu \mu})}{m_{Z'}^2},\nn\\
C_{10}^{\mu \mu}(\rm NP) &=& \bigg[\frac{\pi}{\sqrt{2}G_F \alpha V_{tb}V_{ts}^*}\bigg]\frac{g_L^{bs}(g_L^ {\mu \mu})}{m_{Z'}^2},
\eea 
where $g_L^{bs}$ is the coupling when $b$ quark couple to $s$ quark and $g_L^ {\mu \mu}$ is the $\mu^+-\mu^-$ coupling in the presence of new boson $Z'$ and we have assumed $g_R^ {\mu \mu}=0$. Similarly for $b \to s\nu \bar{\nu}$ transition, the NP contribution arising due to $Z'$ is $C_L^ {\nu \nu} (\rm NP)= C_9^{\mu \mu}(\rm NP)=- C_{10}^{\mu \mu}(\rm NP)$\cite{Buras:2014fpa}. From the neutrino trident production, $g_L ^ {\mu \mu} = 0.5$ has been considered in our paper~\cite{Alok:2017jgr, Alok:2017sui, Huang:2018rys}. The new parameter $g^{bs}_L$ is taken to be real in our analysis.

\section{Numerical analysis and discussions}\label{Num_analysis}

\subsection{Input parameters}
In this section we report all the necessary input parameters used for our computational analysis. 
We consider the masses of mesons, quarks, Fermi coupling constant in the unit of GeV and lifetime of $B_c$ meson in the unit of second, CKM matrix element and fine structure constant from Ref.~\cite{ParticleDataGroup:2020ssz}.
We have adopted the lattice QCD method~\cite{Cooper:2021ofu} and the relativistic quark model~\cite{Ebert:2010dv} based on quasipotential approach for the form factors of $B_c \to D_s$ and $B_c \to D^{\ast}_s$ transitions, respectively. 

The form of the form factors for $B_c \to D_s$ transition in lattice QCD are given as follows

\bea
f(q^2)=P(q^2)^{-1} \sum _{n=0}^{N_n}c^{(n)}\hat{z}^{(n,N_n)},
\eea
where \textcolor{red}{$N_n=3$} and the $q ^2$ dependent pole factor $P(q ^2) = 1-q^2 / M_{res} ^2$ ($M_{res}=5.711$ ($f_0$)~\cite{Lang:2015hza}, 5.4158 ($f_{+,T}$)~\cite{ParticleDataGroup:2020ssz}).  Using the Bourreley-Caprini-Lellouch (BCL) parametrisation~\cite{Bourrely:2008za}, the expressions of $\hat{z} ^{n,N_n}_{0,+,T}$ are given as 

\bea
\hat{z}^{n,N_n}_{0} =z^n, \hspace{0.5cm}  \hat{z}^{n,N_n}_{+,T} = z^n -\frac{n(-1)^{N_n+1-n}}{N_n+1}z^{N_n+1}.
\eea
Here $z(q^2)$ is defined as 
\bea
z(q^2)=\frac{\sqrt{t_+-q^2} - \sqrt{t_+-t_0}}{\sqrt{t_+-q^2} + \sqrt{t_+-t_0}},
\eea
where $t_0=0$ and $t_+ = (m_{B(0^-)}+M_{K(0^-)})^2$ with the masses $M_{B(0^-)}=5.27964$ and $M_{K(0^-)}=0.497611$. The form factor coefficients $c^{(n)}$
are reported in Table \ref{tab_ffs}. For our error analysis, we employ $10\%$ uncertainty in the form factor coefficients $c^{(n)}$.
For all the omitted details we refer to~\cite{Cooper:2021ofu}.

Similarly, the form factors for $B_c \to D_s^*$ transition are defined as

\begin{numcases} {F(q^2) =}
\frac{F(0)}{\Big(1-\frac{q^2}{M^2}\Big)\Big(1-\sigma_1\,\frac{q^2}{M_{B^{\ast}_s}^2} + \sigma_2\,\frac{q^4}{M_{B^{\ast}_s}^4}\Big)}, & for $F=\{V, A_0, T_1\}$\nn\\
\frac{F(0)}{\Big(1-\sigma_1\,\frac{q^2}{M_{B^{\ast}_s}^2} + \sigma_2\,\frac{q^4}{M_{B^{\ast}_s}^4}\Big)}, & for $F=\{A_1, A_2, T_2, T_3\}$
\end{numcases}
Here $M=M_{B_s}$ for $A_0(q^2)$ whereas $M=M_{B_s^*}$ is considered for all other form factors. We use $M_{B_s^*}=5.4254$ GeV from the Ref.~\cite{ParticleDataGroup:2020ssz}. 
The related form factor input parameters for $B_c \to D_s^*$ are reported in the Table \ref{tab_ffs}. 
Similarly, we employ $10\%$ uncertainty in the zero recoil momentum function $F(0)$ for our theoretical error in $B_c \to D_s^*$ form factors.

\begin{table}[htbp]
\centering
\setlength{\tabcolsep}{6pt} 
\renewcommand{\arraystretch}{1} 
\begin{tabular}{|c|c|c|c|c||c|c|c|c|c|c|c|c|}
\hline
$B_c \to\, D_s$ & $ c^{(0)}$ & $c^{(1)}$ &$c^{(2)}$ & $c^{(3)}$ & $B_c \to\, D_s^*$ & $ V$ & $A_0$ &$A_1$ & $A_2$& $T_1$ & $T_2$ & $T_3$\\
\hline
$f_0$ &$0.217$ &$-0.220$ &$1.300$ &$-0.508$ & $F(0)$ &$0.182$ &$0.070$ &$0.089$ &$0.110$ &$0.085$ & $0.085$ & $0.051$ \\
$f_+$ &$0.217$ &$-0.559$ &$5.149$ &$-0.217$ & $\sigma_1$ &$2.133$ &$1.561$ &$2.479$ &$2.833$ &$1.540$ & $2.577$ & $2.783$ \\
$f_T$ &$0.299$ &$-1.501$ &$3.579$ &$-0.348$ & $\sigma_2$ &$1.183$ &$0.192$ &$1.686$ &$2.167$ &$0.248$ & $1.859$ & $2.170$ \\
\hline
\end{tabular}
\caption{\large{The lattice QCD form factor coefficients $c^{(n)}$ for $B_c \to D_s$ transition~\cite{Cooper:2021ofu} and the relativistic quark model form factors 
at $q^2=0$ and the corresponding fitted parameters $\sigma_1$ and $\sigma_2$ for $B_c \to D^{\ast}_s$ transition \cite{Ebert:2010dv}}}
\label{tab_ffs}
\end{table}

\subsection{Fit Results}
To obtain the NP parameter space in the presence of $Z'$ and LQs we perform a naive $\chi ^2$ analysis with the available $b \to s \ell \ell$ experimental data. In the fit we consider specifically the LHCb measurements of five different observables such as $R_K$, $R_{K^{*}}$, $P'_5$, $ BR (B_s \to \phi \mu \mu)$ and $ BR(B_s \to \mu ^+\mu^-)$. Our fit include the latest measurements of $R_K$, $ BR (B_s \to \phi \mu \mu)$ and $ BR(B_s \to \mu ^+\mu^-)$ as reported from LHCb in 2021. For our theoretical computation of the underlying observables we refer to the lattice QCD form factors~\cite{Bouchard:2013eph} for $R_K$ and  the form factors obtained from the combined analysis of LCSR+LQCD for $B \to K^*$ and $B_s \to \phi$ decay processes~\cite{Bharucha:2015bzk}. We define the $\chi ^2$ as
\bea
\chi^2(C_i^{\rm NP})= \sum_i  \frac{\Big ({\cal O}_i^{\rm th}(C_{9,10}^{\mu \mu}(NP)) -{\cal O}_i^{\rm exp} \Big )^2}{(\Delta {\cal O}_i^{\rm exp})^2+(\Delta {\cal O}_i^{\rm sm})^2},
\eea
where ${\cal O}_i ^ {\rm th}$ represent the theoretical expressions including the NP contributions and ${\cal O}_i ^ {\rm exp}$ are the experimental central values. The denominator  includes $1 \sigma$ uncertainties associated with the theoretical and experimental results. From our analysis we obtain the best fit values and the 
corresponding $1\sigma$ range of the NP coupling strengths associated with $Z'$, $S_{1/3}^3$ and $U_{-2/3}^3$ LQs respectively as shown in Table \ref{table-bestfit}. 
The best fit points for the NP coulings of $Z^ \prime$ and LQs are obtained by minimizing the $\chi ^2$ variable.  Similarly, to obtain the allowed $1\sigma$ range of each NP coupling, we impose $\chi ^2 \leq 9.488$ constraint corresponding to 95$\%$ CL. The minimum and maximum value of the $1\sigma$ range are given in Table \ref{table-bestfit}. 

\begin{table}
\centering
\begin{tabular}{|l||*{5}{c|}}\hline
\backslashbox[4cm]{NP models}{Values}
&\makebox[6em]{Best fits}&\makebox[6em]{$1\sigma$ range}
\\\hline\hline
$Z' : g_{bs}^ {\mu \mu}\times 10^{-3}$ &$1.74$&[0.11, 3.60]\\\hline
$S_{1/3}^3 : y_{\ell q}^{\prime \mu b} (y_{\ell q}^{\prime \mu s})^*\times 10^{-4}$ &-8.70&[-15.50, -4.50]\\\hline
$U_{-2/3}^3 : g_{\ell q}^{\prime \mu b} (g_{\ell q}^{\prime \mu s})^*\times 10^{-4}$ &8.70&[4.50, 15.50]\\\hline
\end{tabular}
\caption{\large{The best-fit values and the corresponding $1\sigma$ ranges of the NP couplings associated with $Z^{\prime}$ and LQ models.}}
\label{table-bestfit}
\end{table}
\subsection{Interpretation of $B_c \to D_s^{(*)}(\mu ^+ \mu ^-, \nu \bar{\nu})$ decays in standard model and beyond}

\subsubsection{$B_c \to D_s^{(*)}\mu ^+ \mu ^-$ decays}
\begin{figure}[ht!]
\centering
\includegraphics[scale=0.65]{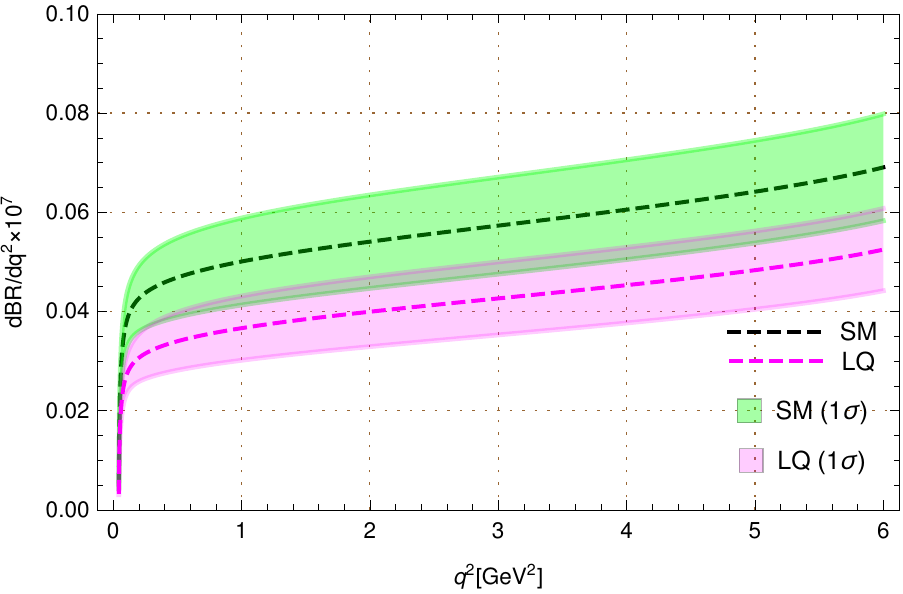} 
\hspace{0.5cm}
\includegraphics[scale=0.65]{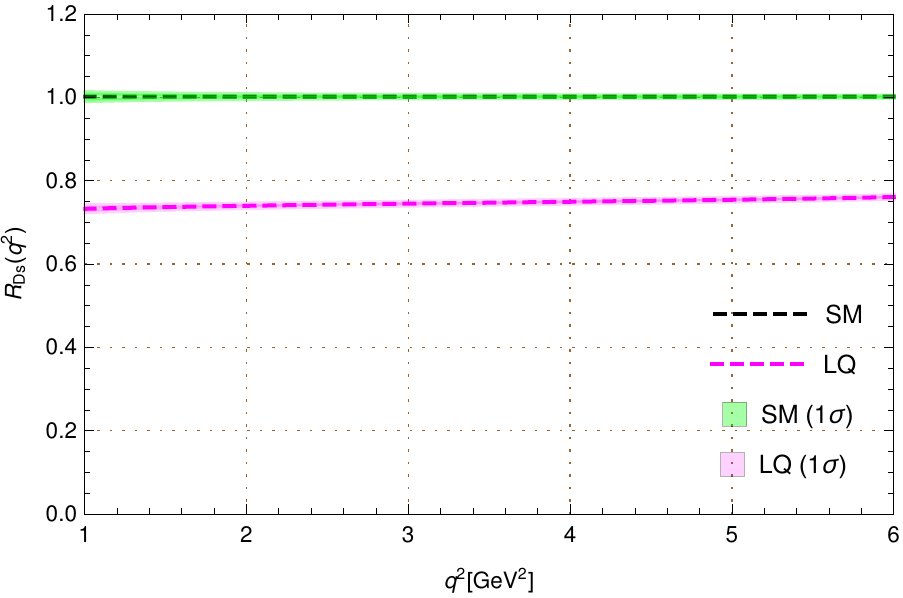}
\vspace{0.2cm}

\includegraphics[scale=0.65]{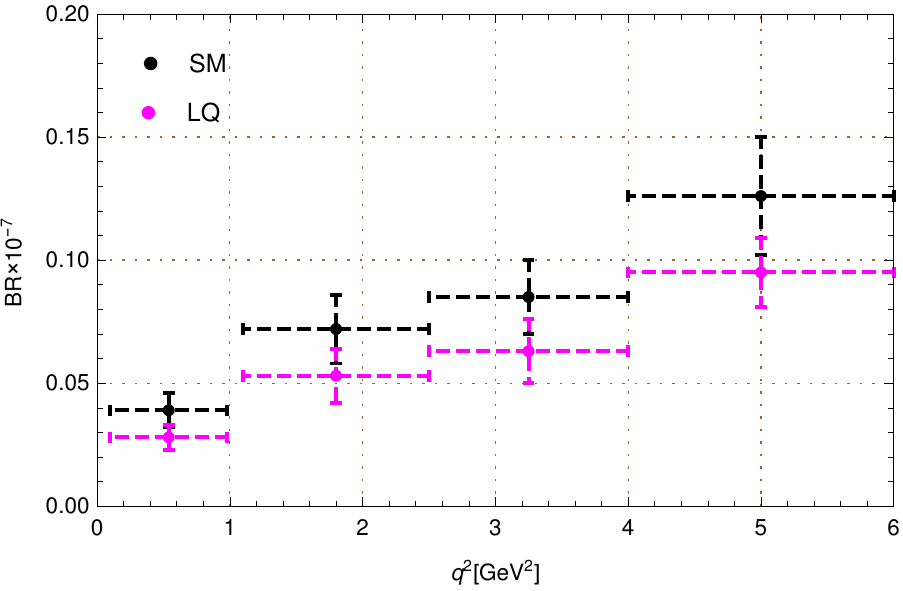} 
\hspace{0.5cm}
\vspace{0.3cm}
\includegraphics[scale=0.65]{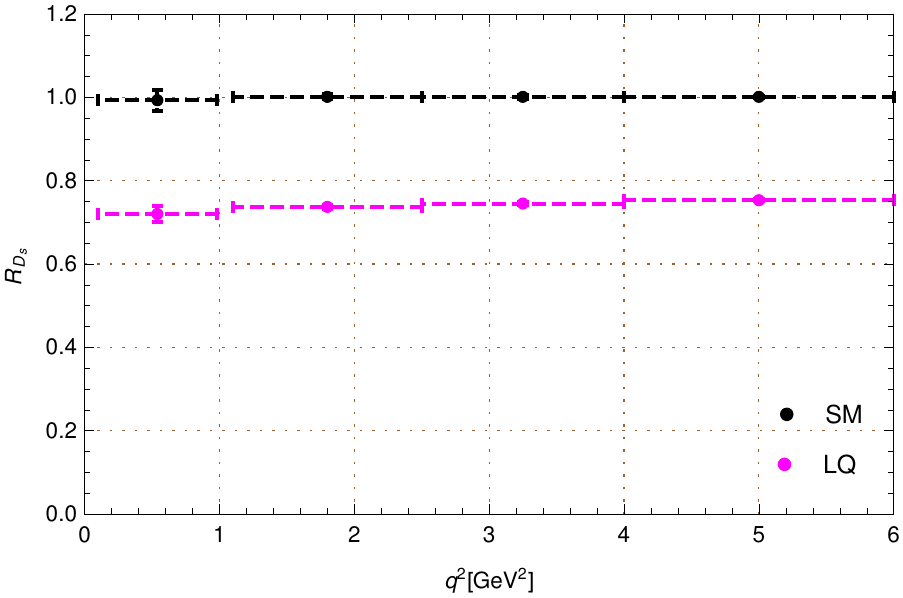}
\caption{\large {The $q^2$ dependency and the bin wise distribution of the branching ratio and the ratio of branching ratio in $B_c \to D_s\,\mu ^+ \mu ^-$
decays in SM and in the presence of $Z'$/LQs}.}
\label{BcDs Fig}
\end{figure}

\begin{figure}[htbp]
\centering
\includegraphics[scale=0.6]{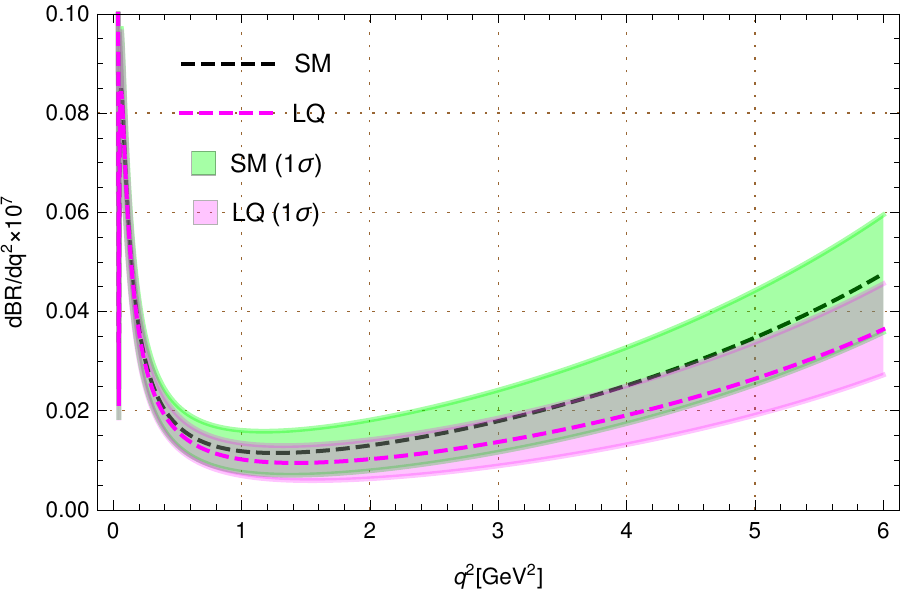} 
\hspace{0.3cm}
\includegraphics[scale=0.6]{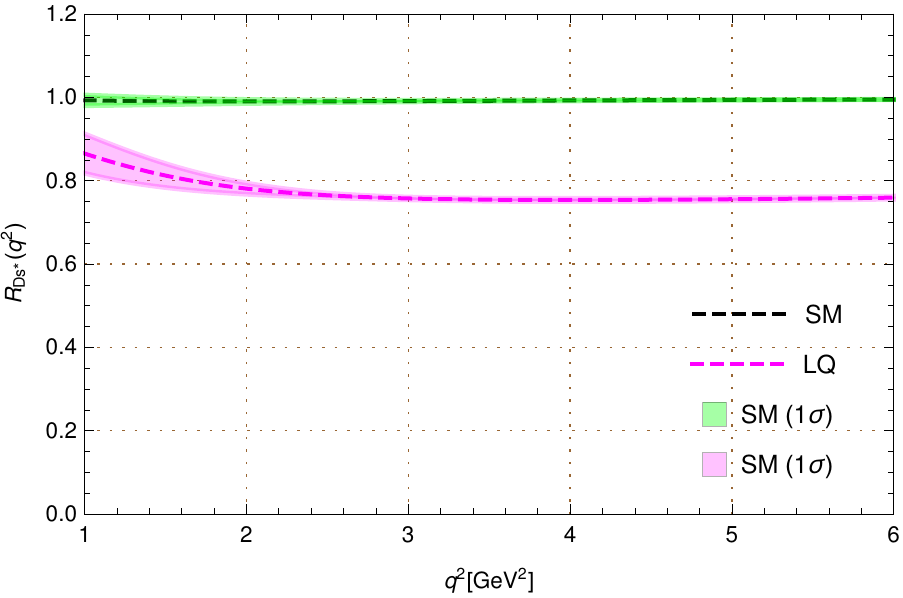}
\hspace{0.3cm}
\includegraphics[scale=0.6]{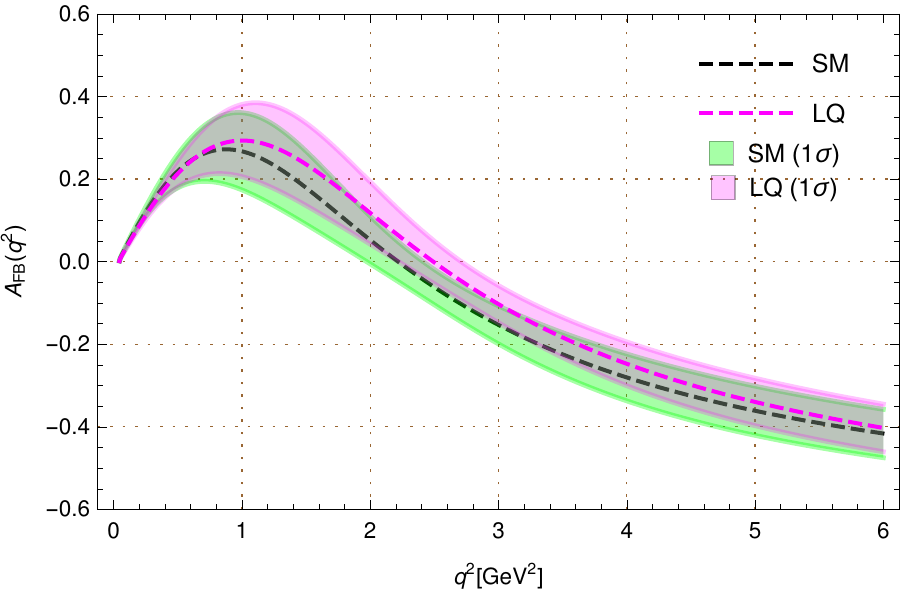}
\vspace{0.2cm}

\includegraphics[scale=0.6]{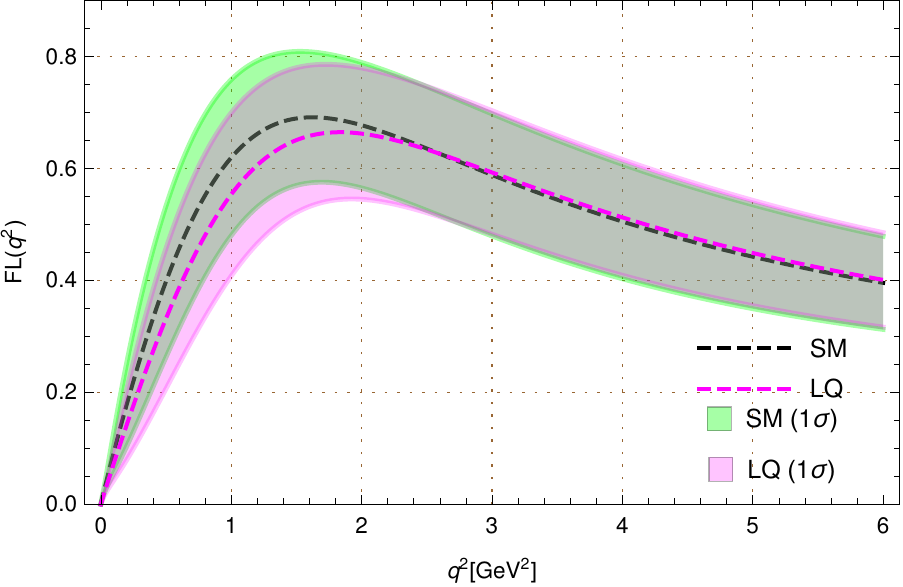} 
\hspace{0.3cm}
\vspace{0.3cm}
\includegraphics[scale=0.6]{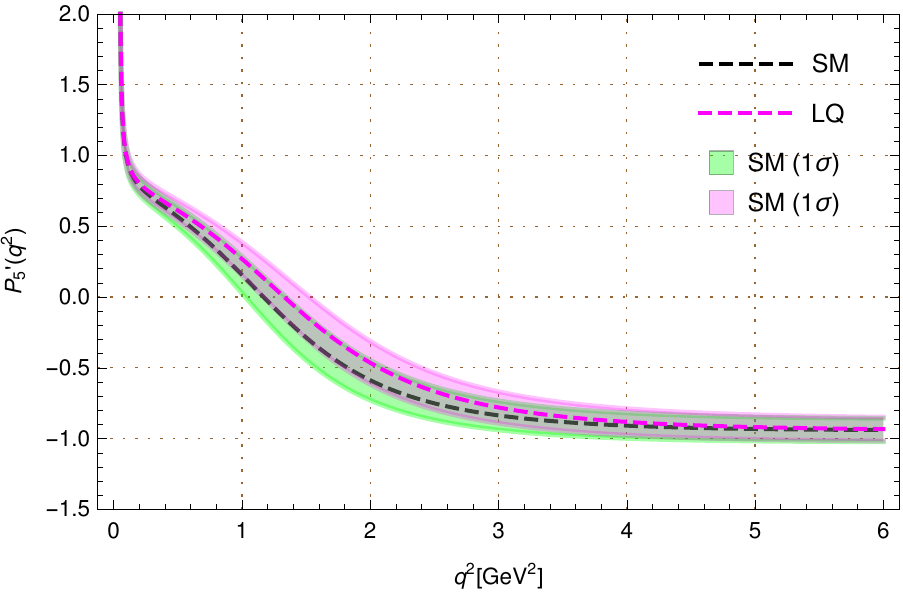}

\includegraphics[scale=0.6]{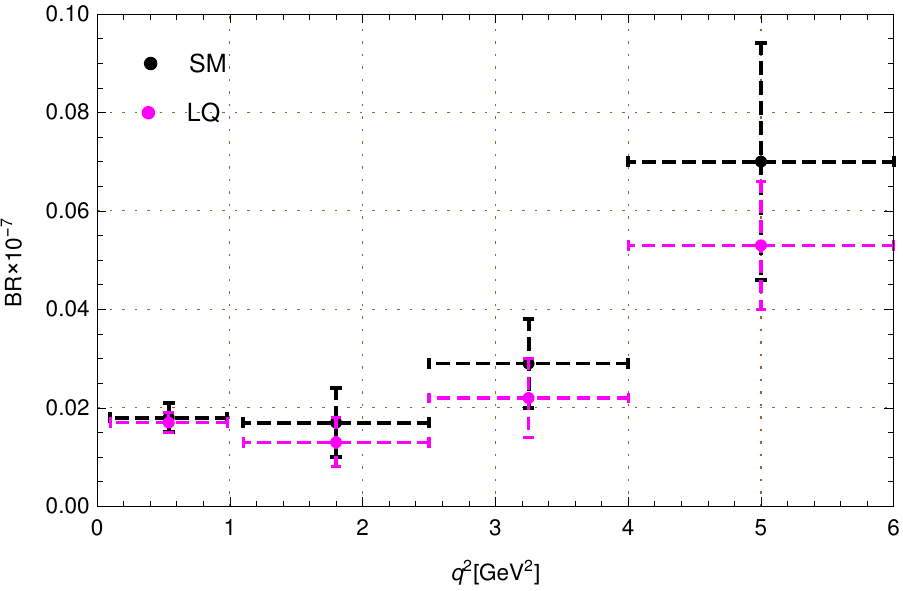}
\hspace{0.3cm} 
\includegraphics[scale=0.6]{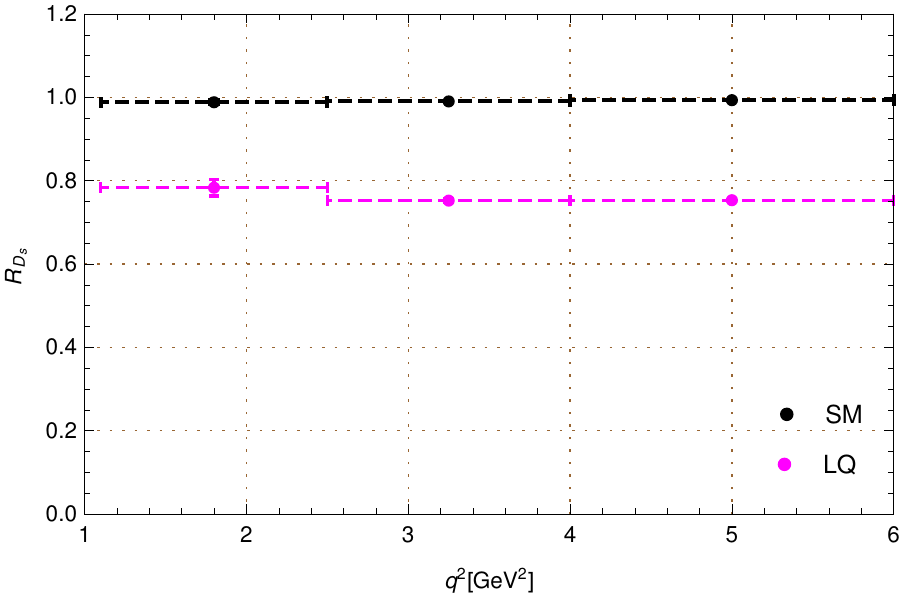}
\hspace{0.3cm}
\includegraphics[scale=0.6]{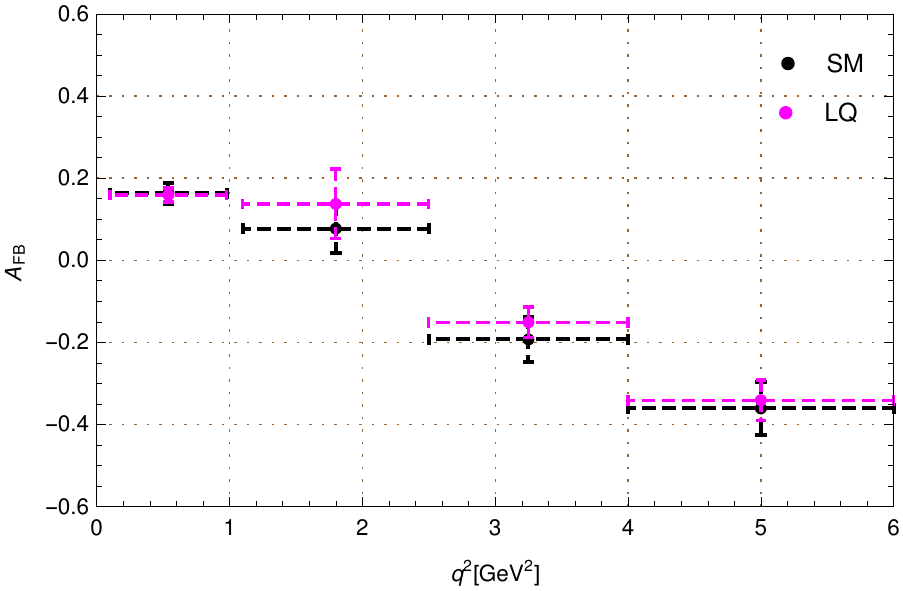} 
\vspace{0.2cm}

\includegraphics[scale=0.6]{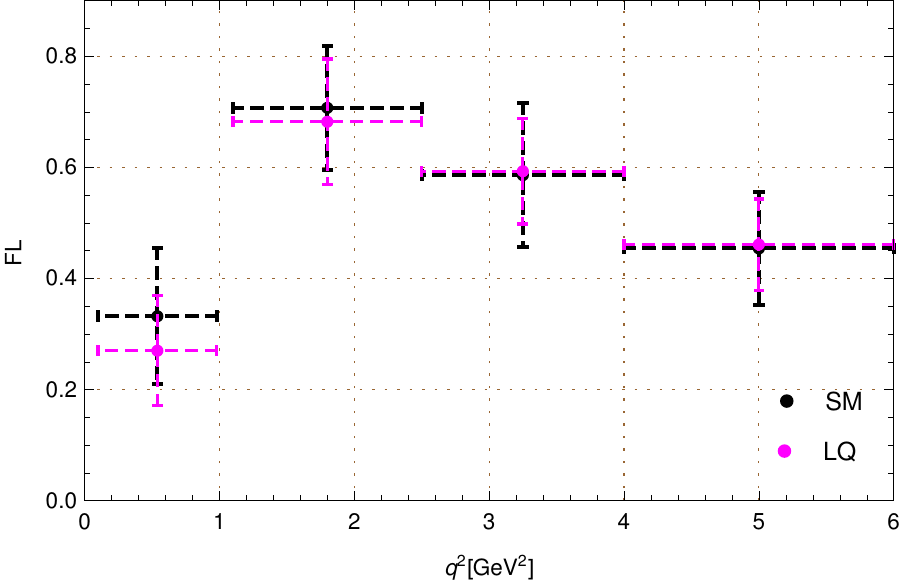} 
\hspace{0.3cm}
\vspace{0.3cm}
\includegraphics[scale=0.6]{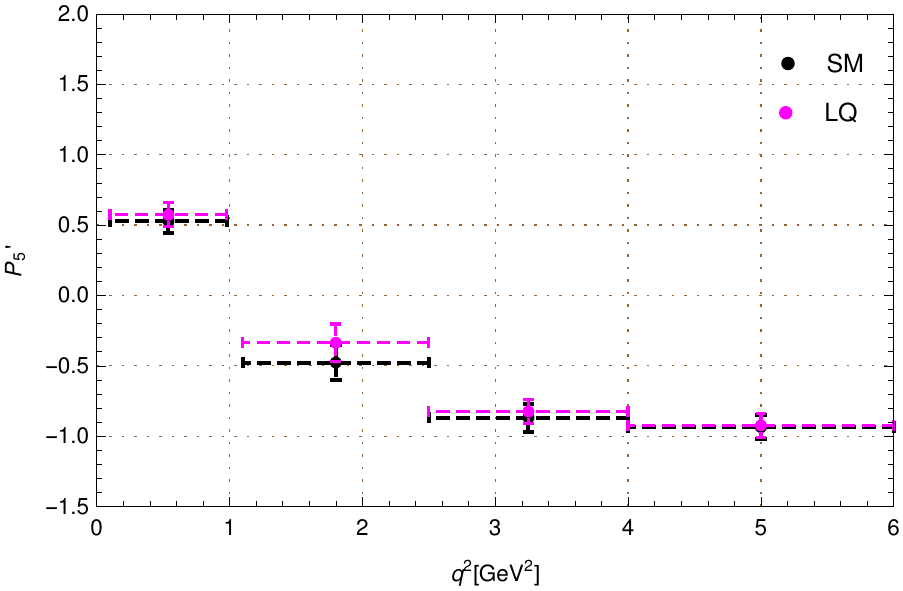} 

\includegraphics[scale=0.6]{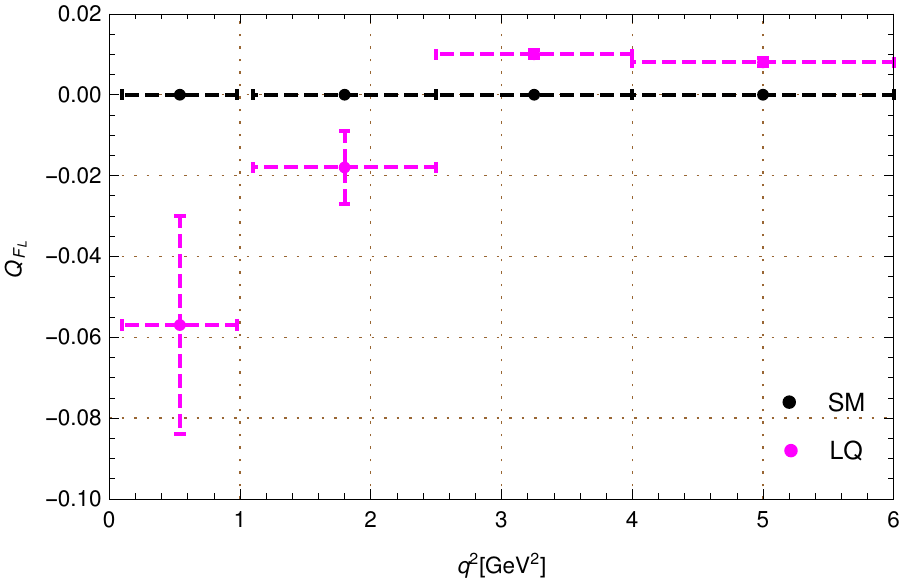} 
\hspace{0.3cm}
\includegraphics[scale=0.6]{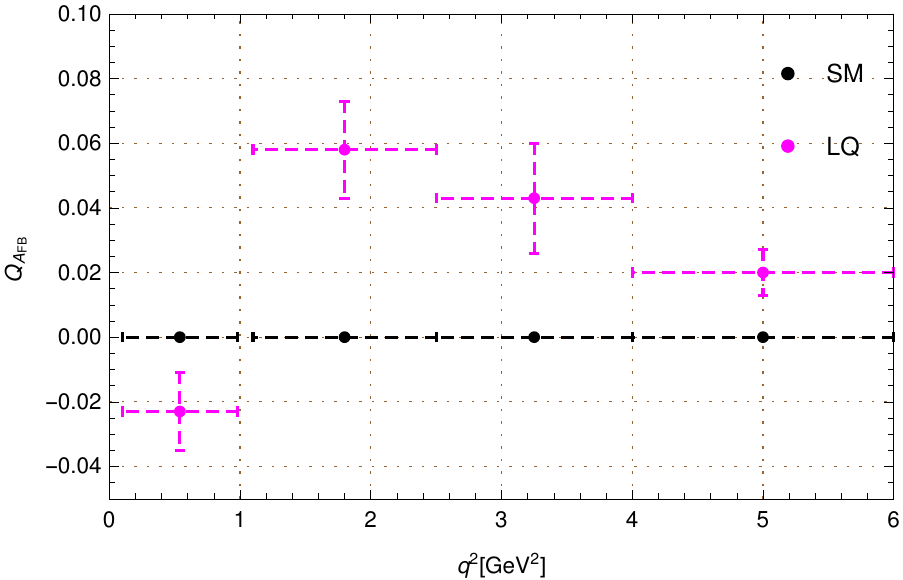} 
\hspace{0.3cm}
\includegraphics[scale=0.6]{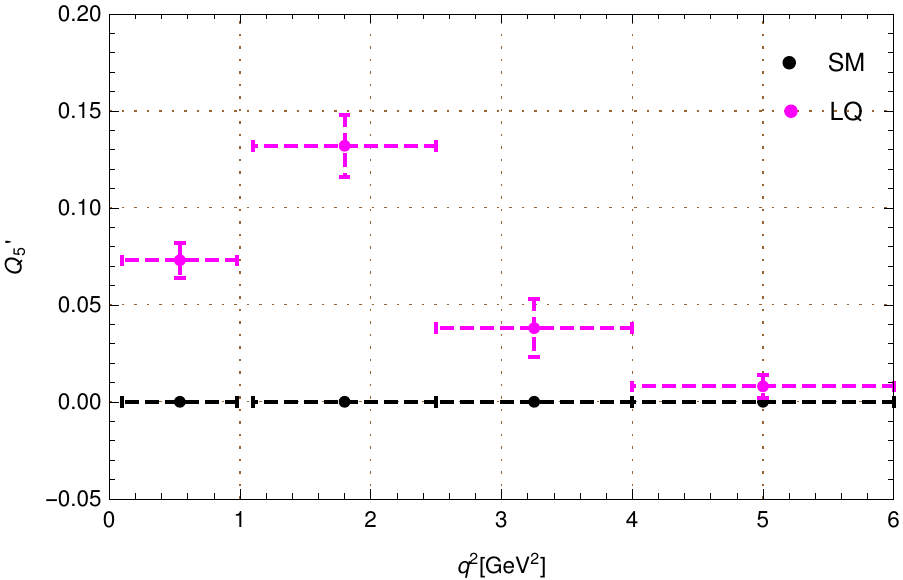} 
\caption{\large {The $q^2$ dependency and the bin wise distribution of various observables such as the differential branching fraction, the ratio of branching ratio 
$R_{D_s^{(*)}}$, the forward backward asymmetry, the lepton polarization asymmetries, the angular observable $P'_5$ and the $Q$ parameters for $B_c \to D_s^{*}\,\mu ^+ \mu ^-$
decays in SM and in the presence of $Z'$/LQs}.}
\label{BDstrar Figures}
\end{figure}

We perform NP studies of $B_c \to D_s^{(*)}\mu ^+ \mu ^-$ decays in the presence of $Z'$, $S_{1/3}^3$ and $U_{-2/3}^3$ LQs which satisfy $C_9 ^{\mu \mu}(NP)=-C_{10} ^{\mu \mu}(NP)$
new physics scenario. Although the NP coupling strengths associated with the $Z'$, $S_{1/3}^3$ and $U_{-2/3}^3$ LQs are different from each other, the 
contribution from the $C_9^{\mu \mu}(NP)=-C_{10}^{\mu \mu}(NP)$ new Wilson coefficients in $b \to s \ell^+ \ell^-$ 
decays are same.
Hence we expect similar NP signature from $Z'$, $S_{1/3}^3$ and $U_{-2/3}^3$ LQs in the underlying $B_c \to D_s^{(*)}\mu ^+ \mu ^-$ decays.
We study various observables such as the differential branching ratio, the forward backward asymmetry, the lepton polarization fraction, the LFU sensitive observables including the ratio of branching ratio $R_{D_s^{(*)}}$ and the difference of the observables associated with $Q$ parameters such as
$Q_{F_L}$, $Q_{A_{FB}}$ and $Q_5^{\prime}$ in the presence of SM as well as new physics. 
In Table~\ref{tab_sm2} we report the central values and the corresponding standard deviation for all the observables in both SM and $Z'$/LQ new physics.
Similarly in Fig.~\ref{BcDs Fig} and Fig.~\ref{BDstrar Figures} we display the corresponding $q^2$ distribution plots as well as $q^2$ integrated bin wise plots for
$B_c \to D_s\mu ^+ \mu ^-$ and $B_c \to D_s^{*}\mu ^+ \mu ^-$ processes respectively.
For the binned plots we choose different bin sizes which are compatible with the LHCb experiments starting from [0.1, 0.98], [1.1, 2.5], [2.5, 4.0], [4.0, 6.0] 
and also [1.1, 6.0].
Similarly, for the $q^2$ distribution plots, we display the central lines and the corresponding $1\sigma$ error band for both SM and $Z'$/LQ new physics scenarios.
The central lines are obtained by considering only the central values of all the input parameters  
and the corresponding $1\sigma$ error bands are obtained by varying the form factors and the CKM matrix element within $1\sigma$.
In SM we obtain the branching fraction to be $\mathcal{O}(10^{-7})$ for $B_c \to D_s^{(*)}\mu ^+ \mu ^-$ decay channels.
The detailed observations of our study are as follows:

\begin{itemize}
\item The $q^2$ dependency of the differential branching fraction for $B_c \to D_s^{(*)}\mu ^+ \mu ^-$ decays are shown in the top - left panel of Fig.~\ref{BcDs Fig}
and Fig.~\ref{BDstrar Figures} respectively. We notice that the differential branching ratio is reduced in the presence of $Z'$/LQ new physics 
and the NP central line lies away from the SM uncertainty band for $B_c \to D_s\,\mu ^+ \mu ^-$ decay. Although the $Z'$/LQ new physics contribution 
in $B_c \to D_s^{*}\mu ^+ \mu ^-$ decay deviate from SM central curve but it cannot be distinguished beyond the SM uncertainty however, slight more deviation
can be found at $q^2>4 \rm GeV^2$. Moreover, partial overlapping of the SM and NP uncertainties can be noticed over the $q^2$. 
Similarly, in the top - right panel of Fig.~\ref{BcDs Fig} and bottom middle panel Fig.~\ref{BDstrar Figures} we display the corresponding binned plots respectively for both the decay modes.
We observe that for $B_c \to D_s\,\mu ^+ \mu ^-$ decay in the all the bins the new physics contribution stand at $>1\sigma$ away from the SM. 
For the decay $B_c \to D_s^{*}\mu ^+ \mu ^-$ however, the NP central values differ from the SM but no such significant observations can be made. 

\item The ratio of branching ratio $R_{D_s^{(*)}}(q^2)$ is constant over the range $q^2\in [0.1, 6.0]$ and is approximately equal to $\sim 1$. 
The uncertainties associated with this observable is almost zero both in SM as well as in the presence of NP contribution. 
The NP contribution from $Z'$/LQ is easily distinguishable from the SM contribution beyond the uncertainties at more than $5\sigma$ as
shown in Fig.~\ref{BcDs Fig} and Fig.~\ref{BDstrar Figures} respectively for both the decays. However the claim of $5 \sigma$ deviation is observed only by considering the best fit points of $Z^{\prime}/ LQ$ coupling strength and neglecting the corresponding experimental error of the measurement. 

\item The $q^2$ distribution of the forward backward asymmetry $A_{FB}(q^2)$ have a zero crossing at $\sim 2.2 \rm GeV^2$ in SM which is different from 
the $Z'$/LQ new physics contribution crossing nearly at $\sim 2.5 \rm GeV^2$ as shown in Fig.~\ref{BDstrar Figures}. Although, there is overlapping between SM 
$Z'$/LQ error bands, however, the new zero crossing point from $Z'$/LQ NP is clearly distinguishable beyond the respective uncertainties.
Similarly, in the binned plots we observe that except for the bin [0.1,0.98], the $A_{FB}$ values are shifted to higher values as compared to the SM estimations 
due to the $Z'$/LQ new physics contribution. However, in fact in all $q^2$ bins the $Z'$/LQ new physics spans less than $1\sigma$ deviation from the SM.

The $Z'$/LQ new physics contribution in the longitudinal polarization fraction $F_L(q^2)$ has shifted from the SM for $q^2< 2 \rm GeV^2$ while in the rest of $q^2$ region
the NP contributions coincides with the SM contribution. No important observations can be drawn from $F_L(q^2)$.

For the angular observable $P_5^{\prime}(q^2)$, 
in the region $q^2 \in [1.1,2.5]$ the $Z'$/LQ new physics contribution can be clearly distinguished from the SM however it lies within the SM error band.
Moreover, the error band corresponding to $Z'$/LQ NP almost overlaps with SM error band and cannot be distinguishable beyond the SM uncertainty.
We do observe the zero crossing for $P_5^{\prime}(q^2)$. In SM we get the zero crossing at $\sim 1.2 \rm GeV^2$ which is different from 
the $Z'$/LQ new physics contribution observed at $\sim 1.4 \rm GeV^2$. However, the zero crossing corresponding to $Z'$/LQ NP cannot be clearly distinguished as it
lies near the overlapping region of both the uncertainties.

\item The observables $\langle Q_{FL} \rangle$, $\langle Q_{A_{FB}} \rangle$ and $\langle Q'_{5} \rangle$ are purely sensitive to test the 
lepton flavor universality violation. The NP contribution in the $Q$ observables can be clearly visualized.
This is because of the reason that all Q's are zeros in SM and hence any non-zero contribution due the NP obviously justifies the beyond SM effects.
From the Fig.~\ref{BDstrar Figures}, for $\langle Q_{FL} \rangle$ we see that the uncertainties associated with the $Z'$/LQ new physics contribution in the lower $q^2$ bins 
such as [0.1, 0.98] and [1.1, 2.5] are huge and hence the deviation reduces nearly to $2\sigma$ whereas, for $q^2>2.5 \rm GeV^2$ the $Z'$/LQ new physics contributions
are clearly distinguishable at more than $5\sigma$.
In the case of $\langle Q_{A_{FB}} \rangle$, the first three bins [0.1, 0.98], [1.1,2.5] and [2.5, 4.0] however show upto $3\sigma$ deviation from the SM, the last bin [4.0, 6.0]
is quite interesting with $>3\sigma$ deviation. 
Similarly, for $\langle Q'_{5} \rangle$ except for the bin [4.0, 6.0] rest of the bins are significantly distinguishable at more than $5\sigma$ from the SM predictions. In all the cases, we have neglected the experimental error.
\end{itemize}

\begin{table}[htbp]
\centering
\setlength{\tabcolsep}{8pt} 
\renewcommand{\arraystretch}{1.5} 
\begin{tabular}{|c|c|c|c|c|c|c|}
\hline
\hline
{Observable} & & {[0.10, 0.98]} & {[1.1, 2.5]} & {[2.5, 4.0]} & {[4.0, 6.0]} & {[1.1, 6.0]} \\
\hline
\multicolumn{7}{|c|}{$B_c \to\, D_s\, \mu^+\,\mu^-$}\\
\hline
\multirow{2}{*}{$BR\times 10^{-7}$}
& $\rm SM$ & $0.039 \pm 0.007$ & $0.072 \pm 0.014$ & $0.085 \pm 0.015$ & $0.126 \pm 0.024$ & $0.284 \pm 0.052$  \\
\cline{2-7}
& $\rm LQ/Z'$ & $0.028\pm 0.005$ & $0.053 \pm 0.011$ & $0.063 \pm 0.013$ & $0.095 \pm 0.014$ & $0.212 \pm 0.036$  \\
\hline 
\multirow{2}{*}{$\langle R_{D_s}^{\mu e} \rangle$}
&$\rm SM$ & $0.993 \pm 0.025$ & $1.001 \pm 0.006$ & $1.001 \pm 0.004$ & $1.001 \pm 0.002$ & $1.001 \pm 0.003$  \\
\cline{2-7}
& $\rm LQ/Z'$ & $0.720\pm 0.020$ & $0.737 \pm 0.005$ & $0.745 \pm 0.004$ & $0.753 \pm 0.003$ &  $0.746 \pm 0.003$ \\
\hline
\multicolumn{7}{|c|}{$B_c \to\, D_s^*\, \mu^+\,\mu^-$}\\
\hline
\multirow{2}{*}{$BR\times 10^{-7}$} 
& $\rm SM$ & $0.018 \pm 0.003$ & $0.017 \pm 0.007$ & $0.029 \pm 0.009$ & $0.070 \pm 0.024$ & $0.116 \pm 0.031$ \\
\cline{2-7}
& $\rm LQ/Z'$ & $0.017 \pm 0.002$ & $0.013 \pm 0.005$ & $0.022 \pm 0.008$ & $0.053 \pm 0.013$ &  $0.088 \pm 0.028$ \\
\hline
\multirow{2}{*}{$\langle F_L \rangle$} 
& $\rm SM$ & $0.332 \pm 0.122$ & $0.707 \pm 0.111$ & $0.586 \pm 0.129$ & $0.454 \pm 0.101$ & $0.525 \pm 0.093$  \\
\cline{2-7}
& $\rm LQ/Z'$ & $0.270 \pm 0.099$ & $0.682 \pm 0.113$ & $0.593 \pm 0.095$ & $0.461\pm 0.082$ & $0.528 \pm 0.101$  \\
\hline
\multirow{2}{*}{$\langle A_{FB} \rangle$} 
&$\rm SM$ & $0.163 \pm 0.026$  & $0.077 \pm 0.060$ & $-0.193 \pm 0.054$ & $-0.361 \pm 0.064$ & $-0.254 \pm 0.051$  \\
\cline{2-7}
&$\rm LQ/Z'$ &  $0.159 \pm 0.017$ & $0.137 \pm 0.085$ & $-0.151 \pm 0.037$ & $-0.341 \pm 0.049$ &  $-0.220 \pm 0.045$ \\
\hline
\multirow{2}{*}{$\langle P_{5}^\prime \rangle$} 
&$\rm SM$ & $0.528 \pm 0.082$  & $-0.477 \pm 0.124$ & $-0.869 \pm 0.100$ & $-0.936 \pm 0.085$ & $-0.842 \pm 0.094$  \\
\cline{2-7}
&$\rm LQ/Z'$ & $0.573 \pm 0.085$  & $-0.337 \pm 0.134$ & $-0.825 \pm 0.084$ & $-0.924 \pm 0.087$ & $-0.803 \pm 0.070$  \\
\hline
\multirow{2}{*}{$\langle R_{D_s^*}^{\mu e} \rangle$} 
&$\rm SM$ & $0.979 \pm 0.011$ & $0.988 \pm 0.005$ & $0.990 \pm 0.001$ & $0.993 \pm 0.000$ & $0.992 \pm 0.001$  \\
\cline{2-7}
& $\rm LQ/Z'$& $0.924 \pm 0.042$ & $0.783 \pm 0.020$ & $0.752 \pm 0.004$ & $0.753\pm 0.004$ & $0.757 \pm 0.003$  \\
\hline
\multirow{1}{*}{$\langle Q_{F_L} \rangle$}
& $\rm LQ/Z'$& $-0.057 \pm 0.027$ & $-0.018 \pm 0.009$ & $0.010 \pm 0.001$ & $0.008\pm 0.001$ & $0.006 \pm 0.001$  \\
\hline
\multirow{1}{*}{$\langle Q_{A_{FB}} \rangle$}
& $\rm LQ/Z'$& $-0.023 \pm 0.012$ & $0.058 \pm 0.015$ & $0.043 \pm 0.017$ & $0.020\pm 0.007$ & $0.0.034 \pm 0.010$  \\
\hline
\multirow{1}{*}{$\langle Q'_{5} \rangle$}
& $\rm LQ/Z'$& $0.073 \pm 0.009$ & $0.132 \pm 0.016$ & $0.038 \pm 0.015$ & $0.008\pm 0.006$ & $0.032 \pm 0.009$  \\
\hline
\end{tabular}
\caption{\large{The SM central value and the corresponding $1\sigma$ standard deviation of various physical observables in SM and in the presence of $Z'$/LQs
for $B_c \to D_s^{(*)}\,\mu ^+ \mu ^-$ decays}}
\label{tab_sm2}
\end{table}

\subsubsection{$B_c \to D_s^{(*)} \nu \bar{\nu}$ decays}
\begin{figure}[htbp]
\centering
\includegraphics[scale=0.65]{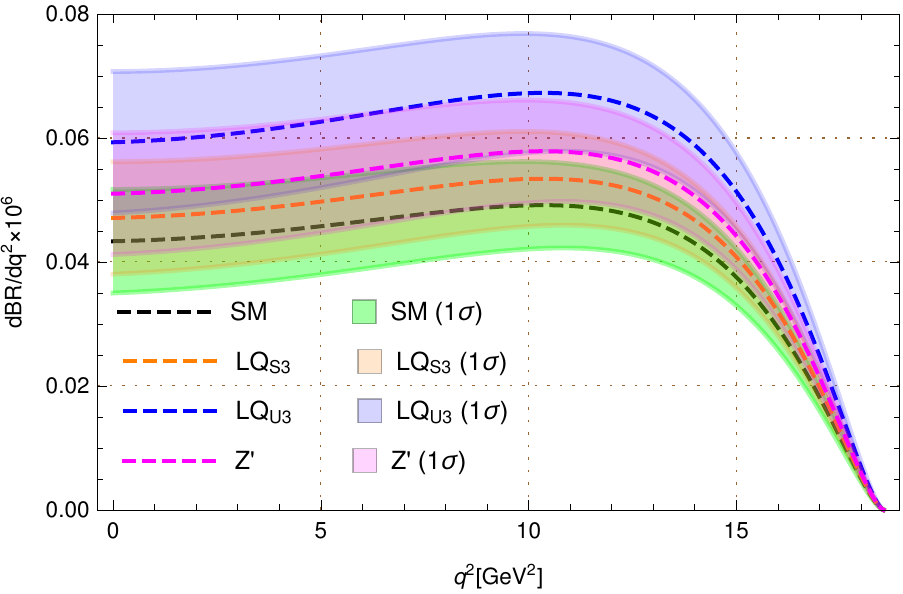} 
\hspace{0.5cm}
\vspace{0.2cm}
\includegraphics[scale=0.65]{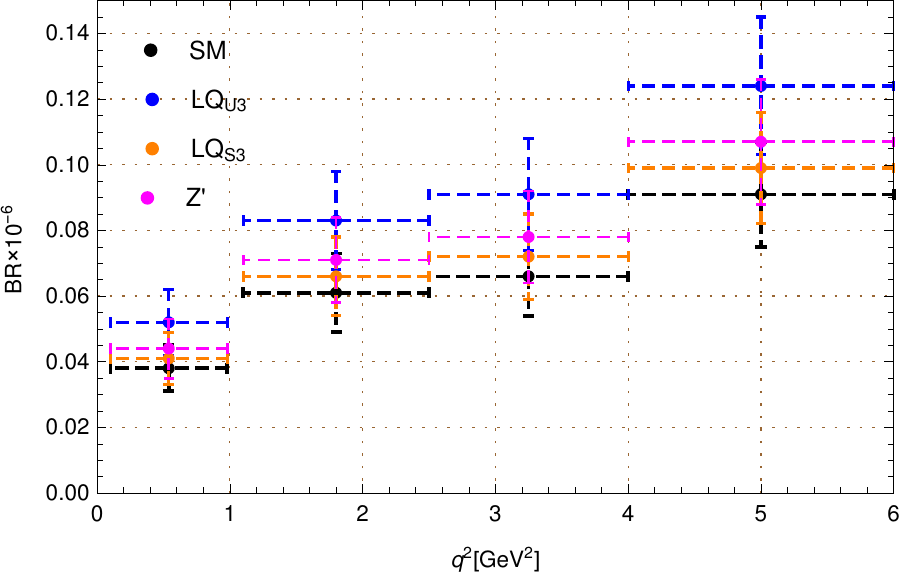} 
\vspace{0.3cm}
\includegraphics[scale=0.65]{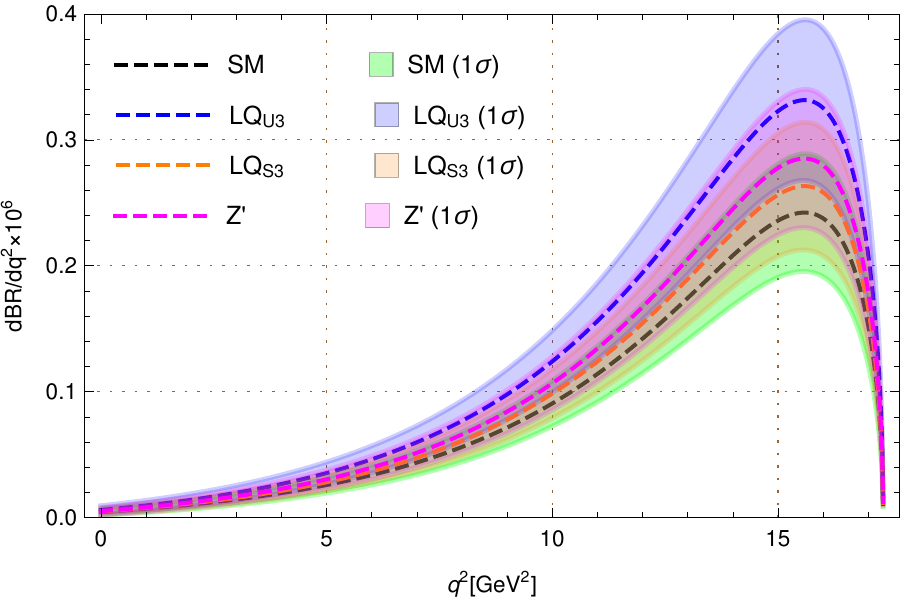} 
\hspace{0.5cm}
\includegraphics[scale=0.65]{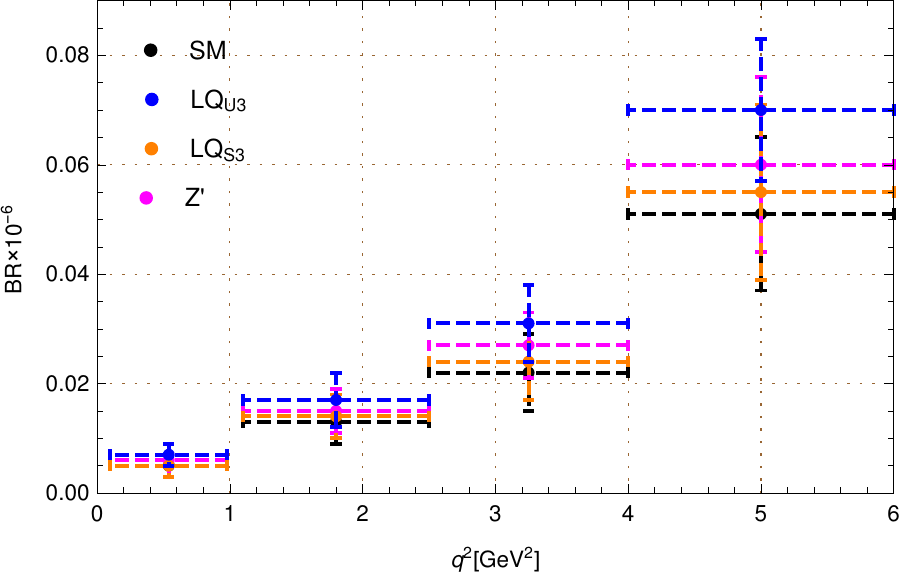}
\hspace{0.5cm}
\includegraphics[scale=0.65]{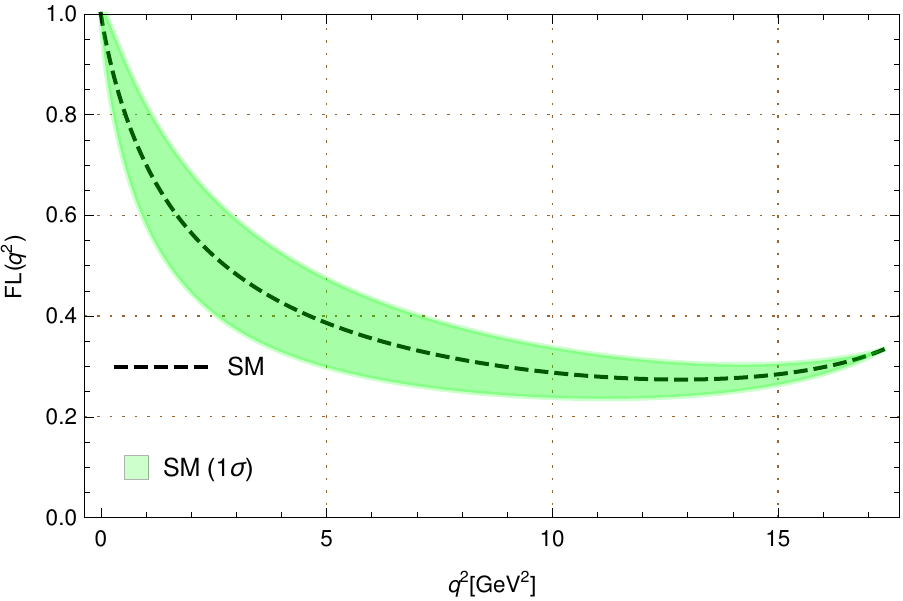} 
\caption{\large{The $q^2$ dependency and bin wise distribution of the branching ratios of $B_c \to D_s \nu \bar{\nu}$ (first row) and $B_c \to D_s^{*} \nu \bar{\nu}$ (second row)
decays in the whole kinematic range in SM and in the presence of $Z'$, $S_{1/3}^3$ and $U_{-2/3}^3$ LQs.
Similarly, the $q^2$ dependency of the lepton polarization fraction of $B_c \to D_s^* \nu \bar{\nu}$ decays in SM is shown in the third row}.}
\label{Bc-Dsnu Figures}
\end{figure}

We know that the neutral semileptonic decays with the neutrinos in the final states are interesting due to the reduced hadronic uncertainties beyond the form factors.
In fact, the $SU(2)_L$ gauge symmetry which treats the charged leptons $(\mu ^+ \mu ^-)$ and neutral leptons $(\nu \bar{\nu})$ to be analogous that invites one to examine
$b\,\to\,s\,\nu \bar{\nu}$ decays in the presence of various beyond the SM scenarios with the implications of available $b \to s \ell^+ \ell^-$ experimental data.
Since we are interested to find out the combined new physics solution which appear both in $b\,\to\,s\,(\ell^+\ell^-,\nu \bar{\nu})$ decays,
we study $B_c \to D_s^{(*)} \nu \bar{\nu}$ decays in SM and also in the presence of $Z'$, $S_{1/3}^3$ and $U_{-2/3}^3$ LQs which satisfy 
$C_9 ^{\mu \mu}(NP)=-C_{10} ^{\mu \mu}(NP)$ scenario in $b \to s \ell^+ \ell^-$ decays. This particular $C_9^{\mu \mu}(NP)=-C_{10}^{\mu \mu}(NP)$ scenario under similar $Z'$/LQ models
have been discussed for the $B_c \to D_s^{(*)}\mu ^+ \mu ^-$ decays in the previous section. 
The new physics contribution to the left handed WC $C_L^{\nu \nu}$ associated with
operator $\mathcal{O}_L^{\nu \nu}$ in $b\,\to\,s\,\nu \bar{\nu}$ decays are related to the corresponding semileptonic WCs such as $C_9 ^{\mu \mu}(NP)$ and $C_{10}^{\mu \mu}(NP)$.
This contribution is different for $Z'$, $S_{1/3}^3$ and $U_{-2/3}^3$ LQs as mentioned in Table~\ref{LQWC}. 
Since we look for the new physics effects associated with left handed neutrinos, the longitudinal polarization fraction appearing in $B_c \to D_s^{*} \nu \bar{\nu}$
decays have no effects beyond the SM however, we only report the SM values for $F_L$ in \ref{tab_sm4}. 
We give predictions for the differential branching fraction in $B_c \to D_s^{(*)} \nu \bar{\nu}$ decays
both in SM and in the presence of several NP models. We obtain the branching fraction for the underlying decays in SM of the $\mathcal{O}(10^{-6})$. 
In Table~\ref{tab_sm3} we report the corresponding branching ratios integrated over different $q^2$ bins in SM and in various NP scenarios.
Similarly, in Fig.~\ref{Bc-Dsnu Figures} we display the $q^2$ dependency of the differential branching ratio in SM, $Z'$, $S_{1/3}^3$ and $U_{-2/3}^3$ LQs.
In the figures we display the central lines and the corresponding $1\sigma$ uncertainty bands respectively for SM and different NP contributions.
The corresponding central lines are obtained by considering the central values of each input parameters and the corresponding $1\sigma$ uncertainty band is obtained by varying the
form factors and CKM matrix element within $1\sigma$. For the different NP models which are constrained by the latest $b \to s \ell^+ \ell^-$ data, we modify the SM WC $C_L^{\nu \nu}$ 
in $b\,\to\,s\,\nu \bar{\nu}$ decays accordingly as reported in Table~\ref{tab_sm2}.
The detailed observations of our study are as follows:

\begin{itemize}
 \item The $q^2$ dependency of $B_c \to D_s^{(*)}\, \nu \bar{\nu}$ decays for the whole kinematic range are displayed in the left panel of Fig.~\ref{Bc-Dsnu Figures}.
 We observe from the plots that the differential branching ratios are enhanced for all the NP contributions and very interestingly the $U_{-2/3}^3$ LQ show significant 
 deviation from the SM curve and lie away from the SM error band. This is because of the reason that the $C_L^{\nu \nu}(NP)-$ the left handed new physics WC in $b\,\to\,s\,\nu \bar{\nu}$ 
 decays for $U_{-2/3}^3$ LQ is rescaled to two times the $C_9^{\mu \mu}(NP)=-C_{10}^{\mu \mu}(NP)$ contribution i.e., $C_L^{\nu \nu}(NP)=2\,C_9^{\mu \mu}(NP)=-2\,C_{10}^{\mu \mu}(NP)$. 
 Similarly, for the $S_{1/3}^3$ LQ the $C_L^{\nu \nu}(NP)=(1/2)C_9^{\mu \mu}(NP)=-(1/2)C_{10}^{\mu \mu}(NP)$.
 On the other hand, for the $Z'$ contribution it is simply $C_L^{\nu \nu}(NP)=C_9^{\mu \mu}(NP)=-C_{10}^{\mu \mu}(NP)$ without any enhancement or reduction. 
 To this end we see that the $S_{1/3}^3$ LQ lie close to the SM whereas $Z'$ contribution lie at the $1\sigma$ boundary of the SM error band. 
Moreover, the error band of our feasible NP model associated with $U_{-2/3}^3$ LQ has distinguishable contributions beyond the uncertainties
 as compared to other NP models such as $S_{1/3}^3$ LQ and $Z^{\prime}$.
 Similarly, on the right panel of Fig.~\ref{Bc-Dsnu Figures} we display the bin wise distribution of the branching ratios only upto $q^2 =6\rm GeV^2$
 with the similar bin sizes as reported earlier. In Table~\ref{tab_sm3} we also have additional bin predictions which are not shown in figure.
 In all the bins we do expect $>1\sigma$ deviation for $U_{-2/3}^3$ LQ, almost $1\sigma$ deviation for $Z'$ and $<1\sigma$ for $S_{1/3}^3$ LQ.
 
\item The interesting fact about the new physics contribution in the longitudinal polarization fraction of $B_c \to D_s^{*}\, \nu \bar{\nu}$ decay is that it is sensitive
 only to the right handed currents. Since in our analysis the new physics
 arising from $Z'$, $S_{1/3}^3$ and $U_{-2/3}^3$ LQs include only the left handed contributions, the polarization fraction will not exhibit any additional new physics effects.
 This is because of the reason that when we define $F_L$ as
 \begin{equation}
  F_L= F_L^{\rm SM}\, \frac{|C_L^{\nu \nu}|^2 + |C_R^{\nu \nu}|^2 - 2 C_L^{\nu \nu} C_R^{\nu \nu}}{|C_L^{\nu \nu}|^2 + |C_R^{\nu \nu}|^2 - \kappa C_L^{\nu \nu} C_R^{\nu \nu}}
 \end{equation}
where, the $C_{L(R)}^{\nu \nu}$ are the Wilson coefficients associated with left (right) handed operators in $b\,\to\,s\,\nu \bar{\nu}$ decays and $\kappa$ is a form factor dependent
parameter~\cite{Altmannshofer:2009ma,Buras:2014fpa}.
In the above equation for $C_R^{\nu \nu}=0$ we obtain $F_L= F_L^{\rm SM}$ and hence any new contribution in $C_L^{\nu \nu}$ will be cancelled.
Therefore in this section we report only the SM predictions for the $F_L$.
The $q^2$ dependency of the longitudinal polarization fraction $F_L(q^2)$ for $B_c \to D_s^{*}\, \nu \bar{\nu}$ decay in the whole kinematic range is displayed in the third row of
 Fig.~\ref{Bc-Dsnu Figures}. In the figure we have shown only the central curve and the corresponding $1\sigma$ error band for SM. Similarly, in the Table~\ref{tab_sm4} we
 report the SM mean and the corresponding standard deviation in various bins including from zero to maximum $q^2$ for which we obtain $F_L=0.301 \pm 0.040$.
\end{itemize}

\begin{table}[htbp]
\centering
\setlength{\tabcolsep}{8pt} 
\renewcommand{\arraystretch}{1.5} 
\begin{tabular}{|c|c|c|c|c|}
\hline
\hline
{$q^2$ bin} & SM & LQ - $\rm S_3$ & LQ - $\rm U_3$ & $Z'$ \\
\hline
\multicolumn{5}{|c|}{$BR (B_c \to D_s \nu \bar{\nu}) \times 10^{-6}$}\\
\hline
\multirow{1}{*}{$[0.1 - 0.98]$} 
& $0.038 \pm 0.007$ & $0.041 \pm 0.008$ & $0.052 \pm 0.010$ & $0.044 \pm 0.009$  \\
\hline
$[1.1 - 2.5]$ & $0.061 \pm 0.012$ & $0.066 \pm 0.012$ & $0.083 \pm 0.015$ & $0.071 \pm 0.013$  \\
\hline
\multirow{1}{*}{$[2.5 - 4.0]$} 
& $0.066 \pm 0.012$ & $0.072 \pm 0.013$ & $0.091 \pm 0.017$ & $0.078 \pm 0.014$ \\
\hline
\multirow{1}{*}{$[4.0 - 6.0]$} 
 & $0.091 \pm 0.016$ & $0.099 \pm 0.017$ & $0.124 \pm 0.021$ & $0.107 \pm 0.019$ \\
\hline
\multirow{1}{*}{$[6.0 - 8.0]$} 
& $0.094 \pm 0.015$  & $0.102 \pm 0.016$ & $0.129 \pm 0.020$ & $0.111 \pm 0.017$ \\
\hline
\multirow{1}{*}{$[11 - 12.5]$} 
& $0.072 \pm 0.010$ & $0.078 \pm 0.010$ & $0.099 \pm 0.013$ & $0.085 \pm 0.011$ \\
\hline
\multirow{1}{*}{$[15 - q^2_{max}]$}
& $ 0.070\pm 0.008$ & $ 0.076\pm 0.009 $ & $0.096 \pm 0.012$ & $0.082 \pm 0.010$ \\
\hline
\multirow{1}{*}{$[1.1 - 6.0]$} 
& $0.219 \pm 0.040$  & $0.238 \pm 0.044$ & $0.299 \pm 0.057$ & $0.257 \pm 0.048$  \\
\hline
\multirow{1}{*}{$[0 - q^2_{max}]$}
& $ 0.758 \pm 0.118$ & $ 0.824 \pm 0.130$ & $1.038 \pm 0.165$ & $0.893 \pm 0.145$ \\
\hline
\multicolumn{5}{|c|}{$BR (B_c \to\,D_s^*\, \nu\,\bar{\nu}) \times 10^{-6}$}\\
\hline
$[0.1 - 0.98]$
& $0.004 \pm 0.002$ & $0.005 \pm 0.002$ & $0.006 \pm 0.003$ & $0.005 \pm 0.002$ \\
\hline
$[1.1 - 2.5]$
& $0.013 \pm 0.004$ & $0.014 \pm 0.004$ & $0.017 \pm 0.005$ & $0.015\pm 0.004$ \\
\hline
$[2.5 - 4.0]$
& $0.022 \pm 0.005$  & $0.024 \pm 0.008$ & $0.031 \pm 0.006$ & $0.026 \pm 0.006$ \\
\hline
$[4.0 - 6.0]$
& $0.051 \pm 0.015$ & $0.055 \pm 0.011$ & $0.070 \pm 0.011$ & $0.060 \pm 0.010$ \\
\hline
$[6.0 - 8.0]$
& $0.0087 \pm 0.020$ & $ 0.094\pm 0.021$ & $ 0.119\pm 0.018$ & $0.102 \pm 0.022$  \\
\hline
$[11 - 12.5]$
& $0.201 \pm 0.045$ & $0.218\pm 0.032$ & $0.275 \pm 0.034$ & $0.236 \pm 0.040$  \\
\hline
$[15 - q^2_{max}]$
& $0.481 \pm 0.085$ & $0.523 \pm 0.092$ & $0.659 \pm 0.089$ & $0.566 \pm 0.100$ \\
\hline
$[1.1 - 6.0]$
& $0.087 \pm 0.022$ & $0.094 \pm 0.025$ & $0.119 \pm 0.021$ & $0.102 \pm 0.020$ \\
\hline
$[0 - q^2_{max}]$
& $1.602 \pm 0.313$ & $1.741 \pm 0.304$ & $2.193 \pm 0.285$ & $1.886 \pm 0.321$ \\                            
\hline
\end{tabular}
\caption{\large{The branching ratios of $B_c \to D_s^{(*)} \nu \bar{\nu}$ decays in different $q^2$ bins in SM and in the presence of $Z'$, $S_{1/3}^3$ and $U_{-2/3}^3$ LQs.}}
\label{tab_sm3}
\end{table} 
\begin{table}[htbp]
\centering
\setlength{\tabcolsep}{6pt} 
\renewcommand{\arraystretch}{1.5} 
\begin{tabular}{|c|c|}
\hline
\hline
{$q^2$ bin} & SM \\
\hline
\multicolumn{2}{|c|}{$F_L (B_c \to D_s^* \nu \bar{\nu})$}\\
\hline
\multirow{1}{*}{$[0.1 - 0.98]$} 
& $0.825 \pm 0.074$  \\
\hline
$[1.1 - 2.5]$ & $0.603 \pm 0.097$  \\
\hline
\multirow{1}{*}{$[2.5 - 4.0]$} 
& $0.475 \pm 0.105$ \\
\hline
\multirow{1}{*}{$[4.0 - 6.0]$} 
 & $0.389 \pm 0.078$ \\
\hline
\multirow{1}{*}{$[6.0 - 8.0]$} 
& $0.333 \pm 0.064$  \\
\hline
\multirow{1}{*}{$[11 - 12.5]$} 
& $0.277 \pm 0.040$ \\
\hline
\multirow{1}{*}{$[15 - q^2_{max}]$}
& $ 0.301 \pm 0.011$ \\
\hline
\multirow{1}{*}{$[1.1 - 6.0]$} 
& $0.444 \pm 0.101$ \\
\hline
\multirow{1}{*}{$[0 - q^2_{max}]$}
& $ 0.301\pm 0.040$ \\
\hline
\end{tabular}
\caption{\large{The lepton polarization fraction of $B_c \to D_s^* \nu \bar{\nu}$ decays in different $q^2$ bins in SM.}}
\label{tab_sm4}
\end{table}

\newpage
\section{Conclusion}\label{conc}

With the experimental data associated with $b \to s \ell^+ \ell^-$ neutral current transition reported in the semileptonic $B \to (K, K^*)\, \mu^+ \mu^-$ 
and $B_s \to \phi \mu^+ \mu^-$ and also purely leptonic $B_s \to \mu^+ \mu^-$ decay processes, we scrutinize the $B_c \to D_s^{(*)}\,\mu ^+ \mu ^-$ 
and $B_c \to D_s^{(*)}\, \nu \bar{\nu}$ decays in the SM followed by the effects in the presence of leptoquark and $Z^ {\prime}$ new physics models. 
Throughout the analysis, we have concentrated on the particular new physics scenario $C_9 ^{\mu \mu}(NP)=-C_{10} ^{\mu \mu}(NP)$ where both the leptoquark and $Z^ {\prime}$ models satisfy
the particular condition.
We obtain the $Z'$ and LQ coupling strengths by fitting the five LHCb experimental data associated with $b \to s \ell^+ \ell^-$ decays including $R_{K^{(*)}}$, $P_5^{\prime}$,
$\mathcal{B}(B_s \to \phi \mu^+ \mu^-)$ and $\mathcal{B}(B_s \to  \mu^+ \mu^-)$. Notably we include the latest updates of $R_K$, $\mathcal{B}(B_s \to \phi \mu^+ \mu^-)$ and 
$\mathcal{B}(B_s \to  \mu^+ \mu^-)$ in our fit analysis.
Interestingly, the new physics analysis pertaining to the $B_c \to D_s$ decay observables by using the lattice QCD form factor results are reported for the first time.

In the decays involving the charged leptons as a final state, we have performed a detailed study of various observables such as the differential branching fraction, 
the forward-backward asymmetry, the lepton polarization asymmetry,
the angular observable $P_5^{\prime}$ and the ratio of branching ratios for $B_c \to D_s^{(*)}\,\mu ^+ \mu ^-$ decays in SM and in the presence of $Z'$/LQ new physics.  
Simultaneously, the similar new physics contributions from $Z'$ and LQs have been inspected in the branching ratios of $B_c \to D_s^{(*)}\nu \bar{\nu}$ decay processes.
We observe from our analysis that the branching ratio is reduced due to $Z'$/LQ in the decays which include the charged leptons as a final state whereas in the processes
involving neutrinos in the final state
the branching ratio is increased for $Z'$, $S_{1/3}^3$ and $U_{-2/3}^3$ LQs. In fact more significant deviation from the SM is found for $U_{-2/3}^3$ particularly in $B_c \to D_s^{(*)}\nu \bar{\nu}$ decays.
Moreover, the zero crossing of the forward-backward asymmetry in $B_c \to D_s^{*} \mu^+ \mu^-$ process is shifted to higher $q^2$ value in the presence of $Z'$/LQ new physics.
Similarly, the LFUV sensitive observables including $R_{D_s^{(*)}}$ and the Q parameters have significant deviations at more than $5\sigma$ from the SM in most of the $q^2$ bins. However the claim of $5\sigma$ deviation can be made only with reference to the best fit point and neglecting the experimental error of the measurement.
In addition, it is important to note that the NP contributions from $Z'$, $S_{1/3}^3$ and $U_{-2/3}^3$ LQs in $B_c\to D_s ^{(*)}\,\mu ^+ \mu ^-$ decays are indistinguishable whereas in the
$B_c\to D_s ^{(*)}\,\nu \bar{\nu}$ case all the three new physics contributions are clearly distinguished from one another.
Having said that the decay modes $B_c \to D_s^{(*)} (\mu^+ \mu^-, \nu \bar{\nu})$ mediated by $b \to s (\ell^+ \ell^-, \nu \bar{\nu})$ transition have received very less attention than the current ongoing
study in $B_{(s)} \to (K,K^*, \phi)\ell^+\ell^-$ processes. 
Hence, the combined study of particular decays $B_c \to D_s^{(*)}\,\mu ^+ \mu ^-$ and $B_c \to D_s^{(*)}\, \nu \bar{\nu}$ will certainly help us in identifying
the possible new physics signatures in both $b \to s \ell^+ \ell^-$ and $b \to s \nu \bar{\nu}$ decays.
Moreover, the improved estimations of the various form factors corresponding to $B_c \to D_s$ and $B_c \to D_s^*$ transitions will be crucial in near future to understand the 
nature of NP. In addition to this, more data samples from the experiments are also required to visualize various observables in $B_c\to D_s^{(*)}\, (\ell^+ \ell^-,\nu \bar{\nu})$ decay processes
and in particular the more experimental studies pertaining to $b \to s \nu \bar{\nu}$ decays can assist to identify the various new physics Lorentz structures.

\acknowledgments 
MKM would like to acknowledge DST INSPIRE fellowship programme for financial support.
NR would like to thank CSIR for the financial help in this work.

\bibliographystyle{ieeetr}
\bibliography{BctoDsstar}

\appendix

\section{Form factors for $B_c \to D_s^{(*)}\, \ell \ell (\ell = e, \mu)$}
The hadronic matrix elements for the exclusive $B_c \to D_s$ transition in terms of form factors is given by \cite{Cooper:2021ofu}
\bea
J_ \mu &=& <D_s|\bar{s}\gamma^{\mu}b|B_c> = f_{+}(q^2)\Big[p_{B_c}^{\mu}+p_{D_s}^{\mu}-\frac{M_{B_c}^2-M_{D_s}^2}{q^2}\,q^{\mu}\Big] + f_0(q^2)
\frac{M_{B_c}^2-M_{D_s}^2}{q^2}\,q^{\mu}\,,\nn \\
J_ \mu ^T &=& <D_s|\bar{s}\sigma^{\mu\nu}\,q_{\nu}b|B_c> =\frac{i\,f_T(q^2)}{M_{B_c}+M_{D_s}}\Big[q^2(p_{B_c}^{\mu}+p_{D_s}^{\mu} - (M_{B_c}^2-
M_{D_s}^2)q^{\mu}\Big]\,,
\eea
where $q =p_{B_c} - p_{D_s}$ and the form factors given above the expression satisfy the following relations:
\bea
f_+(0)= f_0, \hspace*{.5cm} f_0(q^2) = f_+ (q^2) + \frac{q^2}{m_{B_c}^2-m_{D_s}^2}f_-(q^2).
\eea

Similarly, for the $B_c \to D^{\ast}_s$ transition, the hadronic matrix elements can be given in terms of the form factors as~\cite{Ebert:2010dv}
\begin{eqnarray}
<D^{\ast}_s|\bar{s}\gamma^{\mu}b|B_c> &=& \frac{2\,i\,V(q^2)}{M_{B_c}+M_{D^{\ast}_s}}\,\epsilon^{\mu\nu\rho\sigma}\epsilon^{\ast}_{\nu}\,
p_{B_{c_{\rho}}}\,p_{{D^{\ast}_s}_{\sigma}}\,, \nonumber \\
<D^{\ast}_s|\bar{s}\gamma^{\mu}\gamma_5\,b|B_c> &=& 2\,M_{D^{\ast}_s}\,A_0(q^2)\frac{\epsilon^{\ast}\cdot q}{q^2}\,q^{\mu} + 
(M_{B_c} + M_{D^{\ast}_s})\,A_1(q^2)\Big(\epsilon^{{\ast}^{\mu}}-\frac{\epsilon^{\ast}\cdot q}{q^2}\,q^{\mu}\Big)\, \nonumber \\
&&-
A_2(q^2)\frac{\epsilon^{\ast}\cdot q}{M_{B_c} + M_{D^{\ast}_s}}\,\Big[p_{B_c}^{\mu}+p_{D^{\ast}_s}^{\mu}-\frac{M_{B_c}^2-M_{D^{\ast}_s}^2}
{q^2}\,q^{\mu}\Big]\,, \nonumber \\
<D^{\ast}_s|\bar{s}\,i\,\sigma^{\mu\nu}\,q_{\nu}b|B_c> &=& 2\,T_1(q^2)\,\epsilon^{\mu\nu\rho\sigma}\epsilon^{\ast}_{\nu}\,p_{B_{c_{\rho}}}\,
p_{{D^{\ast}_s}_{\sigma}}\,, \nonumber \\
<D^{\ast}_s|\bar{s}\,i\,\sigma^{\mu\nu}\,\gamma_5\,q_{\nu}b|B_c> &=& T_2(q^2)\,\Big[(M_{B_c}^2-M_{D^{\ast}_s}^2)\epsilon^{{\ast}^{\mu}} - 
(\epsilon^{\ast}\cdot q)(p_{B_c}^{\mu}+p_{D^{\ast}_s}^{\mu})\Big] \nonumber \\
&&+ 
T_3(q^2)\,(\epsilon^{\ast}\cdot q)\Big[q^{\mu}-\frac{q^2}
{M_{B_c}^2-M_{D^{\ast}_s}^2}(p_{B_c}^{\mu}+p_{D^{\ast}_s}^{\mu})\Big]\,,
\end{eqnarray}
where $q^{\mu}=(p_B ^ \mu -p_{D_s}^ \mu)$ is the four momentum transfer and $\epsilon_{\mu}$ is polarization vector of the $D^{\ast}_s$ meson. 

\section{Angular coefficients}\label{Ang_coeff}
The $q^2$ dependent angular coefficients required for $B_c \to D_s^{*}\, \ell \ell (\ell = \mu)$ processes are given as follows:
\begin{eqnarray}
I_{1}^{c} &=& \bigg(|A_{L0}|^2 + |A_{R0}|^2\bigg) + 8\frac{m_{l}^2}{q^2} Re\bigg[A_{L0}A_{R0}^{*}\bigg] + 4\frac{m_{l}^2}{q^2}|A_{t}|^2, \nonumber \\
I_{2}^{c} &=& -\beta_{l}^2 \bigg(|A_{L0}|^2 + |A_{R0}|^2\bigg), \nonumber \\
 I_{1}^{s} &=& \frac{3}{4} \bigg[|A_{L\perp}|^2 + |A_{L\parallel}|^2 + |A_{R\perp}|^2 + |A_{R\parallel}|^2\bigg] \bigg(1-\frac{4m_{l}^2}{3q^2}\bigg) +
 \frac{4m_{l}^2}{q^2} Re\bigg[A_{L\perp} A_{R\perp}^{*} + A_{L\parallel} A_{R\parallel}^{*}\bigg], \nonumber \\
 I_{2}^{s} &=& \frac{1}{4} \beta_{l}^2 \bigg[|A_{L\perp}|^2 + |A_{L\parallel}|^2 + |A_{R\perp}|^2 + |A_{R\parallel}|^2\bigg],   \nonumber \\
 I_{3} &=& \frac{1}{2} \beta_{l}^2 \bigg[|A_{L\perp}|^2 - |A_{L\parallel}|^2 + |A_{R\perp}|^2 - |A_{R\parallel}|^2\bigg], \nonumber \\
 I_{4} &=& \frac{1}{\sqrt{2}} \beta_{l}^2 \bigg[Re\bigg(A_{L0}A_{L\parallel}^{*}\bigg) + Re\bigg(A_{R0}A_{R\parallel}^{*}\bigg)\bigg],  \nonumber \\
 I_{5} &=& \sqrt{2} \beta_l \bigg[Re\bigg(A_{L0}A_{L\perp}^{*}\bigg) - Re\bigg(A_{R0}A_{R\perp}^{*}\bigg)\bigg], \nonumber \\
 I_{6} &=& 2 \beta_l \bigg[Re\bigg(A_{L\parallel}A_{L\perp}^{*}\bigg) - Re\bigg(A_{R\parallel}A_{R\perp}^{*}\bigg)\bigg], \nonumber \\
 I_{7} &=& \sqrt{2} \beta_l \bigg[Im\bigg(A_{L0}A_{L\parallel}^{*}\bigg) - Im\bigg(A_{R0}A_{R\parallel}^{*}\bigg)\bigg], \nonumber \\
 I_{8} &=& \frac{1}{\sqrt{2}} \beta_{l}^2 \bigg[Im\bigg(A_{L0}A_{L\perp}^{*}\bigg) + Im\bigg(A_{R0}A_{R\perp}^{*}\bigg)\bigg], \nonumber \\
 I_{9} &=& \beta_{l}^2 \bigg[Im\bigg(A_{L\parallel}A_{L\perp}^{*}\bigg) + Im\bigg(A_{R\parallel}A_{R\perp}^{*}\bigg)\bigg]\,,
\end{eqnarray}
where $\beta _ \ell = \sqrt{1-4m_ \ell ^2/q^2}$. 
According to Ref. \cite{Altmannshofer:2008dz}, the transversity amplitude in terms of form factors and Wilson coefficients are given as
\begin{eqnarray}
 A_{L0} &=& N \frac{1}{2m_{{{D_s^*}}}\sqrt{q^2}}
 \bigg\{(C_{9}^{eff} - C_{10}) \bigg[(m_{B_s}^2 - m_{{D_s^*}}^2 - q^2)(m_{B_s} + m_{{D_s^*}})A_1 - \frac{\lambda}{m_{B_s} + m_{{D_s^*}}}A_2\bigg]+ \nonumber \\  
 && 2\,m_b\, C_{7}^{eff}\, \bigg[(m_{B_s}^2 + 3m_{{D_s^*}}^2 - q^2)T_2 - \frac{\lambda}{m_{B_s}^2 -m_{{D_s^*}}^2}T_3 \bigg] \bigg\}\,, \nonumber \\
 A_{L\perp} &=& - N\sqrt{2} \bigg[(C_{9}^{eff} - C_{10})
 \frac{\sqrt{\lambda}}{m_{B_s} + m_{{D_s^*}}} V + \frac{\sqrt{\lambda}\,2\,m_b\,C_{7}^{eff}}{q^2} T_1 \bigg]\,, \nonumber \\
 A_{L\parallel} &=& N\sqrt{2} \bigg[(C_{9}^{eff} - C_{10})
 (m_{B_s} + m_{{D_s^*}}) A_1 + \frac{2\,m_b\,C_{7}^{eff}(m_{B_s}^2 - m_{{D_s^*}}^2)}{q^2} T_2 \bigg]\,, \nonumber \\
 A_{Lt} &=& N (C_{9}^{eff} - C_{10}) \frac{\sqrt{\lambda}}{\sqrt{q^2}}A_0\,,
\end{eqnarray}
where $\lambda=(m_{B_c}^4\,+m_{D_s}^4\,+q^4\,-\,2\,(m_{B_c}^2m_{D_s}^2+m_{D_s}^2 q^2+q^2 m_{B_c}^2)$ and 
$N$, the normalization constant which is defined as 
\begin{equation}
 N= \bigg[\frac{G_{F}^2\alpha_{em}^2}{3\cdot 2^{10}\pi^5\,m_{B_s}^3}|V_{tb}V_{ts}^{*}|^2 q^2 \sqrt{\lambda}\bigg(1-\frac{4m_{l}^2}{q^2}
 \bigg)^{1/2}\bigg]^{1/2}\,.
\end{equation}
The right chiral component $A_{Ri}$ of the transversity amplitudes can be obtained by replacing $A_{Li}$ by $A_{Li}|_{C_{10} \to -C_{10}} (i= 0, \parallel, \perp, t)$.

\end{document}